\documentclass[12pt, peerreview, onecolumn]{IEEEtran}
\pdfoutput=1
\ifCLASSOPTIONpeerreview

\fi

\usepackage{amssymb, amsmath, graphicx, tikz, amsthm, float, hhline, cite, color, makecell, multirow, subcaption,verbatim}
\usepackage{amsmath,epsfig,amssymb,verbatim,amsopn,color}
\usepackage{cite,xspace}
\usepackage{array,algorithm,algorithmic}
\usepackage{bbm}
\usepackage{array}
\usepackage{multirow}
\usepackage{multicol}
\usepackage{rotating}
\usepackage{dsfont}

\usetikzlibrary{shapes.geometric, arrows}
\tikzstyle{block} = [rectangle, rounded corners, minimum width=3cm, minimum height=1cm,text centered, draw=black]
\tikzstyle{mode} = [diamond, minimum width=3.5cm, minimum height=3.5cm, text centered, draw=black, fill=blue!10]
\tikzstyle{note} = [rectangle, minimum width=2cm, minimum height=1cm, text centered, draw=black, fill=red!20]
\tikzstyle{arrow} = [thick,->,>=stealth]
\usepackage[font=small]{caption}
\newtheorem{prop}{Proposition}
\newcommand{\Tr}[1]{{\color{black}{#1}}}
\begin{document}

\title{Mode Selection, Resource Allocation and Power Control for D2D-Enabled Two-Tier Cellular Network}

\author{Yifei Huang,
~Ali A. Nasir,
~Salman Durrani
~and Xiangyun Zhou

\thanks{Yifei Huang, Salman Durrani and Xiangyun Zhou are with the Research School of Engineering, The Australian National University, Canberra, ACT 2601, Australia. Emails: \{yifei.huang, salman.durrani, xiangyun.zhou\}@anu.edu.au. Ali A. Nasir is with the National University of Sciences and Technology, Islamabad, Pakistan. Email: ali.nasir@seecs.edu.pk.}}

\markboth{IEEE Transactions on Communications submission}{}%

\maketitle
\begin{abstract}
This paper proposes a centralized decision making framework at the macro base station (MBS) for device to device (D2D) communication underlaying a two-tier cellular network. We consider a D2D pair in the presence of an MBS and a femto access point, each serving a user, with quality of service constraints for all users. Our proposed solution encompasses mode selection (choosing between cellular or reuse or dedicated mode), resource allocation (in cellular and dedicated mode) and power control (in reuse mode) within a single framework. The framework prioritizes D2D dedicated mode if the D2D pair are close to each other and orthogonal resources are available. Otherwise, it allows D2D reuse mode if the D2D satisfies both the maximum distance and an additional interference criteria. For reuse mode, we present a geometric vertex search approach to solve the power allocation problem. We analytically prove the validity of this approach and show that it achieves near optimal performance. For cellular and dedicated modes, we show that frequency sharing maximizes sum rate and solve the resource allocation problem in closed form. Our simulations demonstrate the advantages of the proposed framework in terms of the performance gains achieved in D2D mode.
\end{abstract}


\ifCLASSOPTIONonecolumn
\newpage
\else
\fi

\section{Introduction}
\IEEEPARstart{F}{uture} fifth generation (5G) cellular networks are expected to be highly heterogeneous in architecture, with coexistence of macrocells and femtocells as well as device-to-device (D2D) communications~\cite{Wei2014,Asadi2014,Feng2014}. In particular, femtocells are of great importance since they are predicted to generate up to $50\%$ of the voice calls and up to $70\%$ of the mobile data traffic in the near future~\cite{Ekram2013}. The two-tier cellular network architecture, comprising of a central macrocell base station (MBS) and licensed shorter range femtocell access points (FAPs), significantly improves the throughput for in indoor environments as well as the overall network spectrum and energy efficiencies~\cite{Chandrasekhar2009,Ge2014}. However, cross-tier interference needs to be properly managed in such networks~\cite{Ekram2013}.

Recently, D2D communications allowing direct communication between nearby users has been envisaged in 3GPP standards~\cite{Malandrino2014}. The D2D users can utilize unlicensed spectrum (out-of-band) or licensed spectrum (in-band). Compared to out-of-band, in-band D2D can provide more quality of service guarantees~\cite{Asadi2014} and is considered in this paper. In in-band D2D, there are three modes of operation for the D2D users: (i) \textit{dedicated} (or overlay) mode where D2D users are allocated dedicated spectrum, (ii) \textit{reuse} (or underlay) mode where D2D users reuse existing spectrum resources and (iii) \textit{cellular} mode where the D2D users are treated as normal cellular users and relay communications through the MBS. From an operator perspective, determining the type of D2D operation during {\it mode selection} (assuming that neighbor discovery has already been achieved~\cite{Tang2014}) is a crucial initial decision by the network and an important research topic. In dedicated or cellular mode, the fundamental research challenge is resource allocation. In the reuse mode, the fundamental research challenge is interference management via efficient power control. Overall, in order to provide operator managed quality of service guarantees, centralized solutions which have low complexity are desirable.

\textbf{Literature review:} \textit{Mode selection schemes} have been proposed in the literature based on minimum distance between the D2D transmitter (DTx) and D2D receiver (DRx)~\cite{Lin2014}, biased D2D link quality and whether it is at least as good as the cellular uplink quality~\cite{ElSawy2014} or guard zones protecting the MBS~\cite{Doppler2010} or D2D users~\cite{Min2011}. A limitation of the schemes in~\cite{Lin2014,ElSawy2014,Min2011} is that they do not inherently protect the D2D link from interference, while the scheme in~\cite{Doppler2010} does not impose any restrictions on the D2D distance; generally D2D communication is envisaged as short range direct communication. Also in~\cite{Min2011}, the guard zone region surrounding D2D users is primarily used to determine which cellular users are allowed to reuse resources allocated to the D2D users, rather than specifically a mode selection criterion. \Tr{Mode switching and mixed mode approaches, where multiple modes are utilized at once, are also studied in \cite{Feng2015, Tang2016}.}

Once a mode is decided, the network must address \textit{resource allocation} to meet network requirements. In the reuse mode, \textit{power control} is used to manage transmit powers and hence interference. Power control is not guaranteed to provide closed form analytical solutions, but it has been shown that optimal solutions can at least be found from searching from a finite set \cite{Yu2011}, although this claim has only been made with two transmitting sources in the system model. In \cite{Feng2013}, power optimization for one D2D transmitter and one cellular user transmitting during uplink was studied. Since there are two transmitters, the optimization is a simple two-dimensional problem. Power allocation for maximizing sum rate was also studied in \cite{Gjendemsj2008}, where the authors focused on a binary power decision, i.e., powers operate either at their maximums or minimums, and with no signal to interference plus noise ratio (SINR) guarantee for any user. The authors showed that binary power control is optimal for two users, but is suboptimal for arbitrary number of users.
 
Orthogonal resource allocation for D2D was studied in \cite{Yu2011} for both dedicated and cellular modes using the downlink (DL). In each mode, time and frequency allocation was considered, and for each allocation, greedy (unconstrained) and rate constrained optimization was presented. However, in the rate constrained case, only the cellular user has a minimum rate requirement, and thus the possibility exists for the cellular user to be allocated all the resources and leaving the D2D with none. Further, \cite{Yu2011} only considered a single-tier network in its system model. In a two-tier cellular network, a licensed femtocell changes the way resources can be allocated, and in turn changes the maximization of the optimization objective in a non-trivial manner. Note that some papers use joint optimization \cite{Yu2014,Zhong2015,Zhang2015} and/or game theory \cite{Wang2015,Ye2015,Yin2015, Yin2015_2,Zhang2015_2} to solve resource allocation problems. In theory, joint optimization solutions could be optimal, but their complexity often means approximations are required in practice. Further, in two-tier networks, different users may have different constraints or requirements which will further increase the difficulty of finding optimal solutions. Meanwhile, game theory has the advantage that it is a more distributive approach, but does not provide operator managed quality of service guarantees.

Both uplink (UL) and downlink (DL) spectrum resources can be used by in-band D2D. In the literature, there exists works which either use UL~\cite{Feng2013,Yu2014,Song2015} or DL~\cite{Zhu2014,Chen2012}, and also some which consider both \cite{Yu2011,Malandrino2014b}. Generally, interference scenarios are less severe in the UL~\cite{Lin2014_mag,Mumtaz2014,Ye2015}. However, in this paper we assume DL resources are reused as this represents the worst case interference scenarios.\footnote{The methodologies developed in this paper can also be applied to reuse UL resources. We would only need to make a distinction between the two for cellular resource allocation since we assume half-duplex communications.}

In summary, existing works on D2D communications have generally considered mode selection, resource allocation and power control sub-problems either separately or considered a subset of these problems for single and multi-tier cellular networks~\cite{Lin2014,ElSawy2014,Doppler2010,Min2011,Yu2011,Yu2014,Song2015,Zhong2015,Wang2015,Ye2015,Feng2013,Gjendemsj2008,Zhu2014,Chen2012,Malandrino2014b}. To the best of our knowledge, a centralized solution for mode selection, resource allocation and power control in D2D-enabled two-tier cellular networks is still an open problem.


\textbf{Paper Contributions:} In this paper, we propose a base station assisted \textit{D2D decision making framework} (cf. Fig.~2) that incorporates mode selection, resource allocation and power control in a two-tier cellular network. The MBS first decides if D2D dedicated mode is permissible or not based on the DTx-DRx separation distance being small enough and the availability of orthogonal resources. If not, an interference criteria is then applied to determine whether the D2D pair should enter reuse mode or remain in cellular mode. Resource and power allocation is then applied to maximize user sum rates. Compared to joint optimization methods, this multi-stage decision process can arrive at the correct mode and resource allocation in a much more straightforward fashion with less complexity. The major technical contributions of the paper are as follows:
\begin{itemize}
\item We propose a mode selection method that prioritizes D2D dedicated mode if the D2D pair are close to each other and orthogonal resources are available, and otherwise allows reuse mode if the D2D pair satisfies a strict distance and interference criteria. We show that our proposed decision making framework allows \Tr{more} dedicated D2D users than conventional methods, and allows more correct decisions \Tr{(i.e., higher rate)} when resources are shared.

\item For the D2D reuse mode, we (non-trivially) extend the method described in~\cite{Yu2011} to three dimensions to solve the power allocation problem in a two-tier cellular network. In this process, we first analytically prove that (i) sum SINR is quasi-convex in any number of varying powers and (ii) sum rate has the same derivative behaviour as sum SINR (and hence is almost quasi-convex) when one received power dominates in magnitude over others. Then using these results, we propose a simple approach of finding the corners or vertices of the power region to solve the power allocation problem, which achieves near-optimal performance as compared to exhaustive search.

\item For the cellular and D2D dedicated modes, we show that frequency allocation results in higher rates than arbitrary or time sharing resource block allocation. We solve the frequency allocation problem in a two-tier cellular network to maximize the sum rate, while meeting a minimum rate constraint for all the users to ensure fairness. We also present general resource allocation methods, where possible, for arbitrary number of users and transmitters.
\end{itemize}

\textbf{Paper Organization:} This paper is organized as follows. Section~\ref{sec:sys} presents the system model. The proposed framework and mode selection scheme is described in Section~\ref{framework_mode}. The formulation and solution for power allocation problem in reuse mode is presented in Section~\ref{reuse}. The formulation and solution for resource allocation problem in dedicated and cellular modes is presented in Section~\ref{sec:resource}. The results are presented in Section~\ref{simulation}. Finally conclusions are presented in Section~\ref{sec:conc}.

\section{System Model}\label{sec:sys}

\begin{figure}[h]
\includegraphics[width=0.4\textwidth]{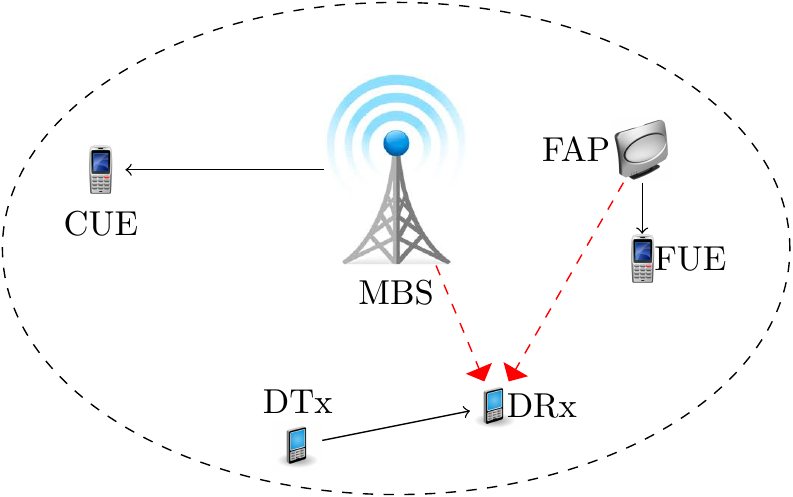}
\centering
\caption{System model comprising of a D2D pair, MBS, FAP, and its served users. Strong interferences to the DRx from the MBS and FAP are shown in red dashed lines.}
\label{fig:system_model}
\end{figure}

We consider a single cell in a two-tier cellular network, as illustrated in Fig.~\ref{fig:system_model}. Our system model is comprised of: (i) an MBS located at the center of the cell, which is serving a single cellular user equipment (CUE), (ii) an FAP serving a single femto user equipment (FUE), and (ii) a D2D pair comprising of a DTx and a DRx located close to each other. All the different user equipments (UEs), MBS and FAP are equipped with single omni-directional antennas. We assume that suitable inter-cell interference control mechanisms, such as fractional frequency reuse, are employed to avoid or manage inter-cell interference~\cite{Novlan2013}. Hence, we study the single cell scenario. Although we study the simplified scenario, as illustrated in Fig.~\ref{fig:system_model}, the proposed framework and resource allocation methods in this paper are applicable to the general scenario with multiple UEs and multiple FAPs, and will be discussed in their respective sections. \Tr{A simple setup was also used in \cite{Ma2016}, with an included discussion on extending the model to more users.}

We assume that the MBS has perfect instantaneous channel state information (CSI) of all the links. This assumption has been widely used in the D2D literature~\cite{Doppler2010, Min2011, Yu2011, Yu2014} and allows benchmark performance to be determined. The mode selection, resource allocation and power control is performed by the MBS in a centralized manner, based on the available perfect CSI. The transmit power of all transmitter nodes is denoted as ${P}_t$ and the maximum transmit power is denoted as ${P}_t^{\textrm{max}}$, where $t \in\{\mathrm{T},\mathrm{M},\mathrm{F}\}$ is the index for the transmitters, and $\mathrm{T}$ denotes DTx, $\mathrm{M}$ denotes MBS and $\mathrm{F}$ denotes FAP. The (minimum) rate at a receiver is denoted as ($\mathcal{R}_r^{\textrm{min}}$) $\mathcal{R}_r$, while the corresponding (minimum) SINR under normalized resource allocation is denoted as ($\gamma_r^{\textrm{min}}$) $\gamma_r$, where $r \in\{\mathrm{\mathrm{R}},\mathrm{\mathrm{C}},\mathrm{E}\}$ is the index for the receivers and $\mathrm{\mathrm{R}}$ denotes DRx, $\mathrm{\mathrm{C}}$ denotes CUE, and $\mathrm{E}$ denotes FUE. All the links are assumed to experience independent block fading.

The instantaneous channel coefficients are composed of small scale fading and large scale path loss denoted as
\begin{equation}
g_{t,r} = h_{t,r} d^{-n}_{t,r}
\end{equation}
where $n$ is the path loss exponent, $h_{t,r}$ is the small scale Rayleigh fading coefficients, which are assumed to be mutually independent and identically distributed (i.i.d.) complex Gaussian random variables with zero mean and unit variance and $d_{t,r}$ denotes the distance in meters between transmitter $t \in\{\mathrm{M},\mathrm{T},\mathrm{F}\}$ and receiver $r \in\{\mathrm{\mathrm{C}},\mathrm{\mathrm{R}},\mathrm{E}\}$. For simplicity, we denote the distance between DTx and DRx $d_{\mathrm{T,R}}$ as simply $d$. All links experience additive white Gaussian noise (AWGN) with power $\sigma^2$.

We use sum rate as our system performance metric with individual maximum power and minimum rate requirements. For clarity, due to the different nature of the D2D modes, we define each problem formulation in their respective sections.

\section{Proposed Framework and Mode selection}\label{framework_mode}

\begin{figure*}[t!]
\centering
\scalebox{0.7}{
\includegraphics[width=1\textwidth]{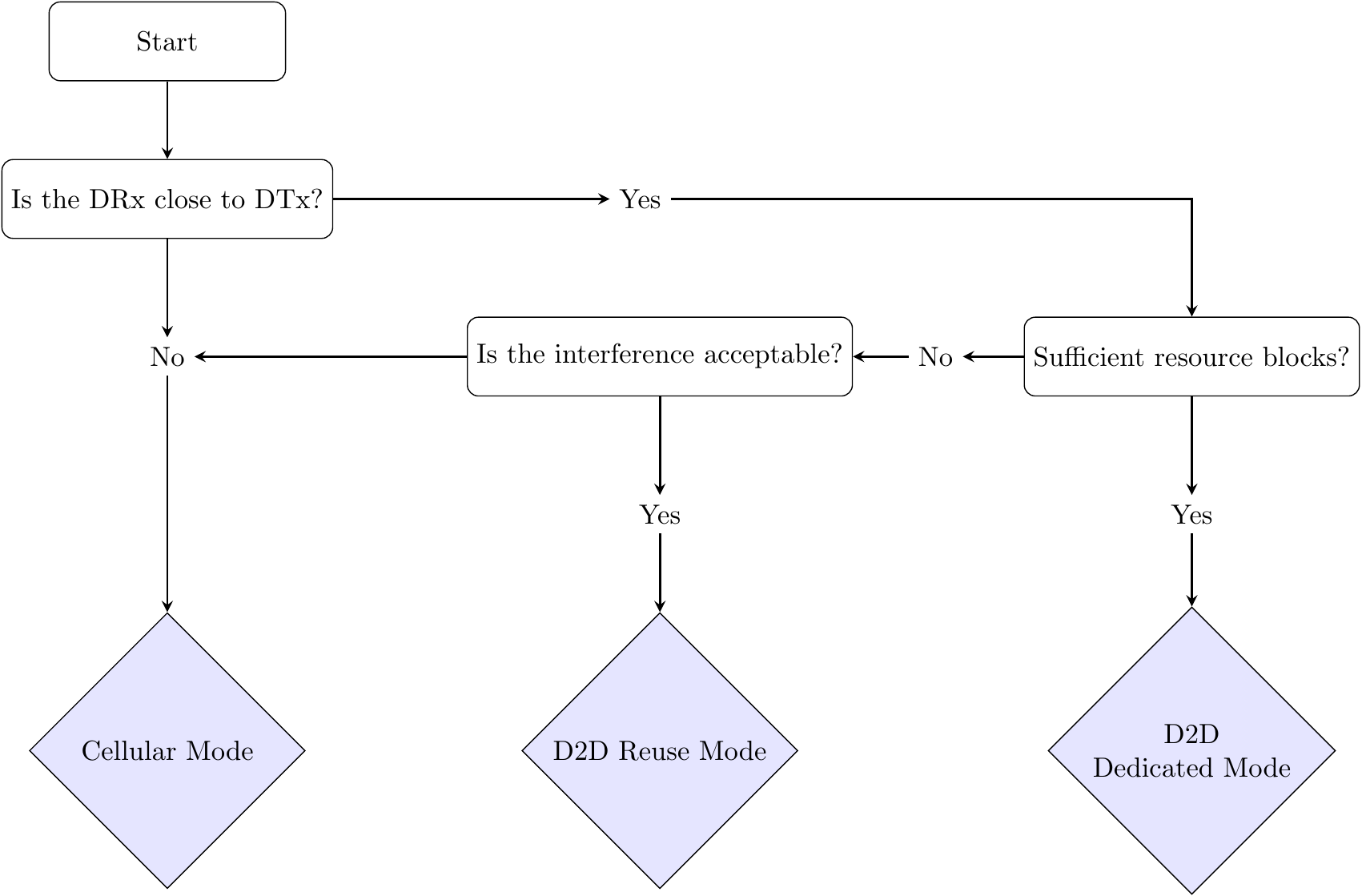}
}
\caption{Proposed MBS assisted \textit{D2D decision making framework} for mode selection, resource allocation and power control in D2D enabled two-tier cellular network.} \label{fig:flowchart}
\end{figure*}

We propose a base station assisted \textit{D2D decision making framework}, as illustrated in Fig.~\ref{fig:flowchart}, to enable the MBS to decide on the correct mode of D2D transmission, and determine the resource parameters of whichever mode is chosen (power for reuse mode, frequency resources for dedicated and cellular modes) that will maximize sum rate subject to maximum transmit power and minimum receiver rate constraints. The main steps in this process are described below:

\begin{enumerate}
\item The MBS first decides whether a potential D2D pair is close enough for D2D communications. Depending on the availability of orthogonal resources and potential interference, dedicated or reuse mode is chosen. Otherwise, the pair remain in cellular mode.

\item If the reuse mode is chosen, then the MBS instructs the CUE, DTx and FAP to control their transmit powers to guarantee quality of service to all receivers. This is done according to the approach proposed in Section \ref{reuse}.

\item If the cellular mode or the dedicated mode is chosen, then the MBS allocates resources to the CUE, FAP and D2D UEs. Since interference is not present, all transmitters can use the maximum transmit power. In the dedicated mode, we assume D2D UEs use the DL resources. In the cellular mode, we assume that both UL and DL resources are used by D2D UEs since the D2D communication is being relayed by the MBS.
\end{enumerate}



\subsection{Mode Selection}

In order to allow as many potential D2D dedicated users as possible, our decision making framework firstly allows a potential D2D pair to enter dedicated mode if they are close enough and orthogonal resources are available. Interference is not considered in this case as using orthogonal resources eliminates interference. If orthogonal resources are not available, the decision to enter reuse mode is then made based on the potential interference. A potential D2D pair can only enter reuse mode if {\it both} of the following criteria are satisfied:
\begin{enumerate}
\item The DRx must be located outside an interference region such that the potential interference is lower than a threshold. Since actual powers are yet to be determined, we assume maximum transmit powers.
\item The distance $d$ between the DTx and DRx must be less than a threshold (should be satisfied from initial step).
\end{enumerate}

To determine the distance and interference thresholds, we recognize that in the cellular mode, the CUE and FUE should always experience better rates compared to D2D due to less interference. Therefore, mode selection is equivalent to finding under what conditions the DRx SINR in cellular mode is better than in D2D mode for the D2D users, i.e.,
\ifCLASSOPTIONonecolumn
\begin{align} \label{eq:MS}
\min\left(\frac{P^{\textrm{max}}_\mathrm{T}g_{\mathrm{T,M}}}{P^{\textrm{max}}_\mathrm{F}g_{\mathrm{F,M}}+\sigma^2}, \frac{P^{\textrm{max}}_\mathrm{M}g_{\mathrm{M,R}}}{P^{\textrm{max}}_\textrm{F}g_{\mathrm{F,R}}+\sigma^2} \right) \geq \frac{P^{\textrm{max}}_\mathrm{T}g_{\mathrm{T,R}}}{P^{\textrm{max}}_\mathrm{M}g_{\mathrm{M,R}}+P^{\textrm{max}}_\mathrm{F}g_{\mathrm{F,R}}+\sigma^2}
\end{align}
\else
\begin{align} \label{eq:MS} \nonumber
& \min\left(\frac{P^{\textrm{max}}_\mathrm{T}g_{\mathrm{T,M}}}{P^{\textrm{max}}_\mathrm{F}g_{\mathrm{F,M}}+\sigma^2}, \frac{P^{\textrm{max}}_\mathrm{M}g_{\mathrm{M,R}}}{P^{\textrm{max}}_Fg_{\mathrm{F,R}}+\sigma^2} \right) \\
& \geq \frac{P^{\textrm{max}}_\mathrm{T}g_{\mathrm{T,R}}}{P^{\textrm{max}}_\mathrm{M}g_{\mathrm{M,R}}+P^{\textrm{max}}_\mathrm{F}g_{\mathrm{F,R}}+\sigma^2}
\end{align}
\fi

\noindent where the $\mathrm{min}(\cdot,\cdot)$ denotes the minimum operator and is used since the rate of the cellular two-hop link is limited by the minimum of the UL and the DL.

Suppose we consider a scenario where a D2D pair is close to each other, but located within a high interference region. From (\ref{eq:MS}), if the interference is greater than a certain threshold
\ifCLASSOPTIONonecolumn
\begin{align}
P^{\textrm{max}}_\mathrm{M}g_{\mathrm{M,R}}+ P^{\textrm{max}}_\mathrm{F}g_{\mathrm{F,R}} \geq P^{\textrm{max}}_\mathrm{T}g_{\mathrm{T,R}} \times
\min\left(\frac{P^{\textrm{max}}_\mathrm{F}g_{\mathrm{F,M}}+\sigma^2}{P^{\textrm{max}}_\mathrm{T}g_{\mathrm{T,M}}}, \frac{P^{\textrm{max}}_\mathrm{F}g_{\mathrm{F,R}}+\sigma^2}{P^{\textrm{max}}_\mathrm{M}g_{\mathrm{M,R}}} \right) -\sigma^2,
\end{align}
\else
\begin{align} \nonumber
& P^{\textrm{max}}_\mathrm{M}g_{\mathrm{M,R}}+ P^{\textrm{max}}_\mathrm{F}g_{\mathrm{F,R}} \geq\\ & P^{\textrm{max}}_\mathrm{T}g_{\mathrm{T,R}} \times
\min\left(\frac{P^{\textrm{max}}_\mathrm{F}g_{\mathrm{F,M}}+\sigma^2}{P^{\textrm{max}}_\mathrm{T}g_{\mathrm{T,M}}}, \frac{P^{\textrm{max}}_\mathrm{F}g_{\mathrm{F,R}}+\sigma^2}{P^{\textrm{max}}_\mathrm{M}g_{\mathrm{M,R}}} \right) -\sigma^2,
\end{align}
\fi

\noindent using D2D mode will be an incorrect decision as it will lead to a lower rate. Note that this threshold value is a conservative estimate as it does not consider the overall system improvement from the CUE or FUE. When the correct mode selection decision is made, the rate achieved by the D2D pair, and also the overall system, will be greater.

A similar argument can be made for a second scenario where the DTx and DRx are located outside a high-interference region, but are far apart. Rearranging (\ref{eq:MS}) to solve for the D2D separation distance, we find that D2D mode would be an incorrect decision if the DRx is outside the interference region, but the D2D pair is separated by a distance of
\ifCLASSOPTIONonecolumn
\begin{align} \label{eq:distance_threshold}
d_\textrm{adaptive} \geq \sqrt[n]{\frac{h_{\mathrm{T,R}}P^{\textrm{max}}_\mathrm{D}}{(P^{\textrm{max}}_\mathrm{M}g_{\mathrm{M,R}}+P^{\textrm{max}}_\mathrm{F}g_{\mathrm{F,R}}+\sigma^2)}} \times
\sqrt[n]{ \min\left(\frac{P^{\textrm{max}}_\mathrm{F}g_{\mathrm{F,M}}+\sigma^2}{P^{\textrm{max}}_\mathrm{D}g_{\mathrm{T,M}}}, \frac{P^{\textrm{max}}_\mathrm{F}g_{\mathrm{F,R}}+\sigma^2}{P^{\textrm{max}}_\mathrm{M}g_{\mathrm{M,R}}} \right)} .
\end{align}
\else
\begin{align} \label{eq:distance_threshold} \nonumber
& d_\textrm{adaptive} \geq \sqrt[n]{\frac{h_{\mathrm{T,R}}P^{\textrm{max}}_\mathrm{D}}{(P^{\textrm{max}}_\mathrm{M}g_{\mathrm{M,R}}+P^{\textrm{max}}_\mathrm{F}g_{\mathrm{F,R}}+\sigma^2)}} \times \\
&\sqrt[n]{ \min\left(\frac{P^{\textrm{max}}_\mathrm{F}g_{\mathrm{F,M}}+\sigma^2}{P^{\textrm{max}}_\mathrm{D}g_{\mathrm{T,M}}}, \frac{P^{\textrm{max}}_\mathrm{F}g_{\mathrm{F,R}}+\sigma^2}{P^{\textrm{max}}_\mathrm{M}g_{\mathrm{M,R}}} \right)} .
\end{align}
\fi

\noindent where $n$ is the path loss index.

However, when the DRx is close to an interference source, \eqref{eq:distance_threshold} may provide an unnecessarily small threshold, and therefore limit the number of D2D pairs. Thus, in our framework we chose the maximum threshold between \eqref{eq:distance_threshold} and a predetermined value $d_\textrm{constant}$, i.e.,
\begin{equation}
d \leq \max\{d_\textrm{constant}, d_\textrm{adaptive}\} \label{eq:dist_new}
\end{equation}
The benefits of this approach will be illustrated using simulation results in Section~\ref{results:modeselec}.

\section{Power Allocation in Reuse Mode} \label{reuse}

In this section, we solve the overall sum throughput optimization problem in the reuse mode. In reuse mode, the problem reduces to finding the optimal powers that can maximize the sum throughput objective while meeting individual minimum rate requirements. We extend the method in~\cite{Yu2011} for the case of two transmitters to three transmitters to solve the power allocation problem in a two-tier cellular network. We present a geometric representation of the problem for the case of three transmitters (i.e., DTx, MBS and FAP) and present a near-optimal\footnote{\Tr{ We use the expression ``near optimal" to describe the closeness of the solution to the optimal solution, rather than to define it as a specific solution or class of solutions.}} solution approaching that of exhaustive search.

\subsection{Problem Formulation}

Our overall system aim is to maximize the sum rate with individual transmit power and receiver rate constraints. We can formulate the optimization problem as follows:
\ifCLASSOPTIONonecolumn
\begin{align} \label{eq:reuse_sum_rate}
\begin{split}
\underset{P_\mathrm{T},P_\mathrm{M},P_\mathrm{F}}{\text{max}}& \left\{ \mathcal{R} \triangleq
\log_2\left(1+\frac{P_\mathrm{T}g_{\mathrm{T},\mathrm{R}}}{P_\mathrm{M}g_{\mathrm{M,R}} + P_\mathrm{F}g_{\mathrm{F,R}}+\sigma^2}\right) +\log_2\left(1+\frac{P_\mathrm{M}g_{\mathrm{M,C}}}{P_\mathrm{T}g_{\mathrm{T},\mathrm{C}}+P_\mathrm{F}g_{\mathrm{F,C}}+\sigma^2}\right) \right.
\\ & \left.+ \log_2\left(1+\frac{P_\mathrm{F}g_{\mathrm{F,E}}}{P_\mathrm{T}g_{\mathrm{T,E}} + P_\mathrm{M}g_{\mathrm{M,E}}+\sigma^2}\right)
\right\}
\end{split}
\end{align}
\else
\begin{align} \label{eq:reuse_sum_rate}
\begin{split}
& \underset{P_\mathrm{T},P_\mathrm{M},P_\mathrm{F}}{\text{max}} \left\{ \mathcal{R} \triangleq
\log_2\left(1+\frac{P_\mathrm{T}g_{\mathrm{T},\mathrm{R}}}{P_\mathrm{M}g_{\mathrm{M,R}} + P_\mathrm{F}g_{\mathrm{F,R}}+\sigma^2}\right) \right. \\ & \left.+\log_2\left(1+\frac{P_\mathrm{M}g_{\mathrm{M,C}}}{P_\mathrm{T}g_{\mathrm{T},\mathrm{C}}+P_\mathrm{F}g_{\mathrm{F,C}}+\sigma^2}\right) \right.
\\ & \left.+ \log_2\left(1+\frac{P_\mathrm{F}g_{\mathrm{F,E}}}{P_\mathrm{T}g_{\mathrm{T,E}} + P_\mathrm{M}g_{\mathrm{M,E}}+\sigma^2}\right)
\right\}
\end{split}
\end{align}
\fi
such that
\begin{subequations}
\begin{align}
& P_t \leq P^{\textrm{max}}_t, t \in\{\mathrm{M},\mathrm{T},\mathrm{F}\} \label{eq:P_max}\\
& \frac{P_\mathrm{T}g_{\mathrm{T},\mathrm{R}}}{P_\mathrm{M}g_{\mathrm{M,R}} + P_\mathrm{F}g_{\mathrm{F,R}}+\sigma^2} \geq \gamma_\mathrm{R}^{\textrm{min}}\label{eq:gamma_D2D}\\
& \frac{P_\mathrm{M}g_{\mathrm{M,C}}}{P_\mathrm{T}g_{\mathrm{T},\mathrm{C}}+P_\mathrm{F}g_{\mathrm{F,C}}+\sigma^2} \geq \gamma_\mathrm{C}^{\textrm{min}} \label{eq:gamma_CUE}\\
& \frac{P_\mathrm{F}g_{\mathrm{F,E}}}{P_\mathrm{T}g_{\mathrm{T,E}} + P_\mathrm{M}g_{\mathrm{M,E}}+\sigma^2} \geq \gamma_{\mathrm{E}}^{\textrm{min}} \label{eq:gamma_FUE}
\end{align}
\end{subequations}

\noindent where \eqref{eq:P_max} represents the maximum power constraints for each transmitter, while \eqref{eq:gamma_D2D}$-$\eqref{eq:gamma_FUE} are minimum SINR requirements. Note that for reuse mode, since all resources are shared and allocation is not considered, a minimum rate constraint is equivalent to a minimum SINR constraint.

\subsection{Geometric Representation}

We adopt a geometric approach to determine the optimal powers. To graphically represent the admissible powers, we first set orthogonal axes to be the powers. Next, setting constraints \eqref{eq:gamma_D2D}$-$\eqref{eq:gamma_FUE} to equality and rearranging, we obtain
\begin{subequations}
\begin{align}
\label{eq:f1} & f_\mathrm{T} \triangleq g_{\mathrm{T},\mathrm{R}}P_\mathrm{T} - \gamma^{\textrm{min}}_\mathrm{R}g_{\mathrm{M,R}}P_\mathrm{M} - \gamma^{\textrm{min}}_\mathrm{R}g_{\mathrm{F,R}}P_\mathrm{F}-\gamma^{\textrm{min}}_\mathrm{R}\sigma^2 = 0, \\
\label{eq:f2} & f_\mathrm{M} \triangleq -\gamma^{\textrm{min}}_\mathrm{C}g_{\mathrm{T},\mathrm{C}}P_\mathrm{T} + g_{\mathrm{M,C}}P_\mathrm{M} - \gamma^{\textrm{min}}_\mathrm{C}g_{\mathrm{F,C}} P_\mathrm{F} - \gamma^{\textrm{min}}_\mathrm{C}\sigma^2 = 0, \\
\label{eq:f3} & f_\mathrm{F} \triangleq -\gamma^{\textrm{min}}_\mathrm{E}g_{\mathrm{T,E}}P_\mathrm{T} - \gamma^{\textrm{min}}_\mathrm{E}g_{\mathrm{M,E}}P_\mathrm{M} + g_\mathrm{F,E}P_\mathrm{F} - \gamma^{\textrm{min}}_\mathrm{E}\sigma^2 = 0,
\end{align}
\end{subequations} which represent planes in 3-dimensional space. The planes themselves represent the relationship between each node's power and the SINR thresholds. Each plane focuses on one threshold, and thus we refer to \eqref{eq:f1}$-$\eqref{eq:f3} as the D2D, MBS, and FAP planes respectively. Each plane intersects with its respective axis at their respective minimum powers $P^{\textrm{min}}_t$. Note that while the thresholds are stated in terms of the receiving node of that link, the powers are of the transmitting node.

We can plot \eqref{eq:f1}$-$\eqref{eq:f3} using their inequalities to obtain a $3$-dimensional upper right corner region within a cube\footnote{Strictly speaking, the region is a rectangular prism, but for conciseness we will use `cube' to describe this region.}, the faces of which represent the maximum individual power constraints. The top right corner of this cube has the maximum power coordinates $(P^{\textrm{max}}_\mathrm{T}, P^{\textrm{max}}_\mathrm{M}, P^{\textrm{max}}_\mathrm{F})$.

The smallest possible transmit powers, $P^{\mathrm{min}}_t$, that satisfy each users' SINR requirement can be calculated from (\ref{eq:gamma_D2D})$-$(\ref{eq:gamma_FUE}) when there is no interference from the other transmissions. Therefore, the range of admissible powers is
\begin{subequations}
\begin{align}
P^{\mathrm{min}}_\mathrm{T} = \frac{\gamma^{\mathrm{min}}_\mathrm{R}\sigma^2}{g_{\mathrm{T},\mathrm{R}}} \leq P_\mathrm{T} \leq P^{\textrm{max}}_\mathrm{T},\\
P^{\mathrm{min}}_\mathrm{M} = \frac{\gamma^{\mathrm{min}}_\mathrm{C}\sigma^2}{g_{\mathrm{M,C}}} \leq P_\mathrm{M} \leq P^{\textrm{max}}_\mathrm{M},\\
P^{\mathrm{min}}_\mathrm{F} = \frac{\gamma^{\mathrm{min}}_{\mathrm{E}}\sigma^2}{g_{\mathrm{F,E}}} \leq P_\mathrm{F} \leq P^{\textrm{max}}_\mathrm{F}.
\end{align}
\end{subequations}

Meanwhile, the minimum powers that jointly satisfy the individual user rate constraints can be found by simultaneously solving \eqref{eq:f1}$-$\eqref{eq:f3} using standard methods such as Cramer's rule. Note that these powers will not maximize sum rate.

We assume that the coefficient matrix formed from \eqref{eq:f1}$-$\eqref{eq:f3} is full rank, i.e., the three planes intersect at a point $Q$, whose coordinates are all positive values since they represent physical transmission powers. We find that the reuse mode is a viable option only if each signal strength is relatively large compared to the interference, making it easier to satisfy SINR constraints. This conclusion is consistent with others in the literature~\cite{Gamage2014}.

\begin{figure*}[t!]
\centering
\begin{subfigure}[t]{0.3\textwidth}
\includegraphics[width=1\textwidth]{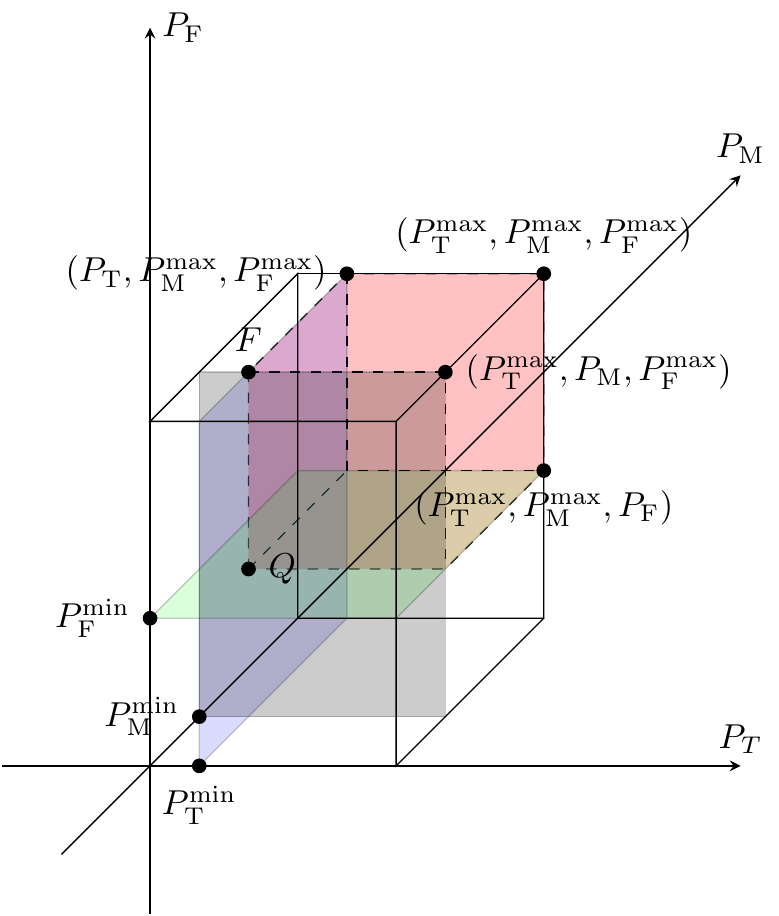}
\caption{All thresholds are satisfied.}
\label{fig:all_thresholds}
\end{subfigure}
~~~
\begin{subfigure}[t]{0.3\textwidth}
\includegraphics[width=1\textwidth]{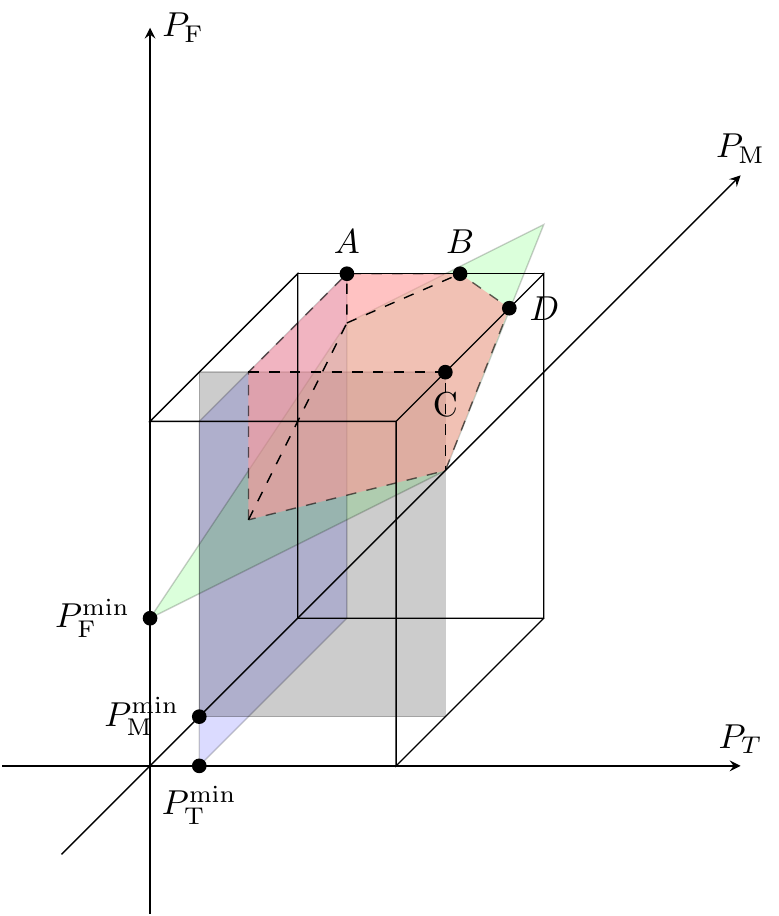}
\caption{Two thresholds are satisfied.}
\label{fig:two_thresholds}
\end{subfigure}
~~~
\begin{subfigure}[t]{0.3\textwidth}
\includegraphics[width=1\textwidth]{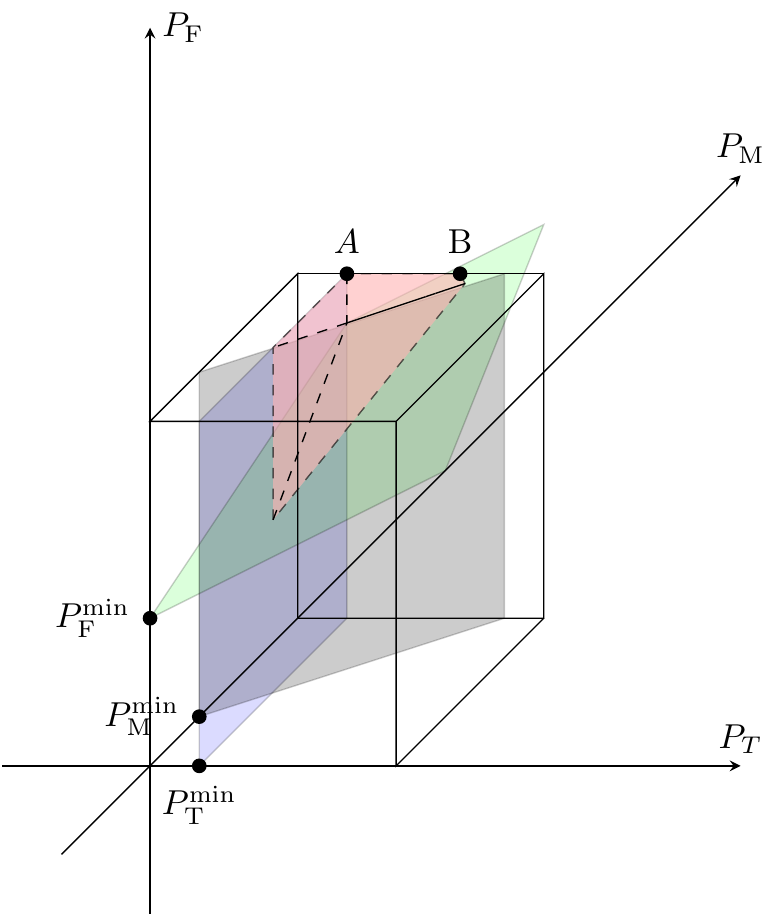}
\caption{One threshold is satisfied.}
\label{fig:one_threshold}
\end{subfigure}
\caption{Illustration of different types of power regions.}
\vspace{-2mm}
\end{figure*}

The admissible {\it power region} is formed by the intersection of the three planes in $3$-dimensional space, and is bounded by these three planes and the three faces of the cube. The optimal powers lie within this power region. In order to avoid an exhaustive search, which would be computationally expensive, we propose a near optimal solution which reduces the process to testing and selecting the optimal powers from a finite set.

\subsection{Proposed Solution - Vertex Search}
In this paper, we adopt the simple approach of finding the corners or vertices of the power region to test for the optimal powers. This approach relies on the following two mathematical conditions:
\begin{enumerate}
\item The optimal powers cannot lie in the interior of the power region, and must be on a boundary.
\item The objective function is quasi-convex on a boundary, ensuring that the maximum values are at the endpoints/vertices.
\end{enumerate}

The first condition was in fact proved in~\cite{Gjendemsj2008}, and thus it is known that at least one of the powers is at its maximum when maximizing sum rate. However, this only states that the optimal solutions exist on the \textit{boundary} of the power region, which includes vertices as well as higher dimensional edges and faces that contain an infinite number of points. Thus, this conclusion from~\cite{Gjendemsj2008} alone is not sufficient to obtain the finite set of points which will give the optimal solution. For two transmitters, it has been proven that the optimal power lies on the corners or vertices of the power region~\cite{Feng2013, Yu2011}, a fact that relies on the convexity of the sum rate function for two powers. However, it is well known that in general, the sum rate expression in \eqref{eq:reuse_sum_rate} is non-convex with respect to arbitrary combinations of varying powers. Consequently, for arbitrary number of transmitters, the optimal powers may not necessarily lie on the vertices of the power region, leading to a possibly infinite set of points to test.

To prove the second condition and justify searching the vertices to maximize sum rate for arbitrary number of powers, we present the following two propositions.
\begin{prop}\label{prop:1}
Sum SINR is a quasi-convex function for any combination of varying powers. Hence, it is also jointly quasi-convex in all powers.
\end{prop}
\begin{IEEEproof}
See Appendix \ref{appendix:SINR_quasiconvex}.
\end{IEEEproof}
\theoremstyle{remark}
\newtheorem{remark}{Remark}
\begin{remark}
Since sum SINR is a quasi-convex function, the powers maximizing it will lie on the one of the vertices of the power region.
\end{remark}
\begin{prop}\label{prop:2}
When one receive power dominates, global maxima and minima for sum rate and sum SINR will occur at the same locations.
\end{prop}
\begin{IEEEproof}
We prove in Appendix \ref{appendix:asymptotic} that when one receive power dominates, e.g., an order of magnitude larger than others, sum SINR in \eqref{eq:sum_SINR} and the inner $\log$ term in \eqref{eq:sum_rate} have the same {\it asymptotic} derivatives, meaning that the two functions will `follow' each other more and more closely the larger the dominant power is. Since logarithm is a monotonic function and does not change the locations of local maxima or minima, this implies that the same powers that maximize sum SINR will also maximize sum rate.
\end{IEEEproof}
\begin{remark}
Since global maxima of sum SINR will be at the vertices of the power region, the same vertices will also give near-optimal solutions for sum rate.
\end{remark}

Note that approximations such as reformulating the objective function as a geometric program (GP) \cite{Chiang2007} can be used to solve \eqref{eq:reuse_sum_rate}. However, we show in the results section that our proposed simple approach yields near optimal solutions quite close to those obtained using exhaustive search and GP, but does not require an iterative approach.

In the next subsection, we present a systematic way of obtaining the coordinates of the vertices of the power region by solving relevant sets of SINR equations.
\begin{table*}
\footnotesize
\caption{Finite set of vertices (suboptimal powers) for reuse mode.}
\begin{center}
\begin{tabular}{ | c| c | c |c| }
\hline
\bf{Type} & \bf{Condition}& \bf{\makecell{Number \\ of points}} & \bf{Set of vertices (suboptimal powers)}\\ \hhline{|=|=|=|=|}
Face point & All & 9 &\makecell{ \\$(\{P_\mathrm{T}, P_\mathrm{M}\}|^{\{f_\mathrm{T},f_\mathrm{M}\}}, P^{\textrm{max}}_\mathrm{F}),(\{P_\mathrm{T}, P_\mathrm{M}\}|^{\{f_\mathrm{T},f_\mathrm{F}\}}, P^{\textrm{max}}_\mathrm{F})$, \\\\ $(\{P_\mathrm{T}, P_\mathrm{M}\}|^{\{f_\mathrm{M},f_\mathrm{F}\}}, P^{\textrm{max}}_\mathrm{F})$, $(\{P_\mathrm{T}, P_\mathrm{F}\}|^{\{f_\mathrm{T},f_\mathrm{M}\}}, P^{\textrm{max}}_\mathrm{M})$, \\\\$(\{P_\mathrm{T}, P_\mathrm{F}\}|^{\{f_\mathrm{T},f_\mathrm{F}\}}, P^{\textrm{max}}_\mathrm{M})$, $(\{P_\mathrm{T}, P_\mathrm{F}\}|^{\{f_\mathrm{M},f_\mathrm{F}\}}, P^{\textrm{max}}_\mathrm{M})$,\\\\ $( P^{\textrm{max}}_\mathrm{T}, \{P_\mathrm{M},P_\mathrm{F}\}|^{\{f_\mathrm{T},f_\mathrm{M}\}})$,$( P^{\textrm{max}}_\mathrm{T}, \{P_\mathrm{M},P_\mathrm{F}\}|^{\{f_\mathrm{T},f_\mathrm{F}\}})$,\\\\$( P^{\textrm{max}}_\mathrm{T}, \{P_\mathrm{M},P_\mathrm{F}\}|^{\{f_\mathrm{M},f_\mathrm{F}\}})$ } \\ \hline
\makecell{Edge point -\\ All thresholds \\ satisfied} & \makecell{$\gamma'_\mathrm{R} \geq \gamma^{\textrm{min}}_\mathrm{R}, \gamma'_\mathrm{C} \geq \gamma^{\textrm{min}}_\mathrm{C} $\\\\ $\text{ and } \gamma'_{\mathrm{E}} \geq \gamma^{\textrm{min}}_{\mathrm{E}}$} & 4 & \makecell{\\ $(P_\mathrm{T}|^{\{f_\mathrm{T}\}}, P^{\textrm{max}}_\mathrm{M}, P^{\textrm{max}}_\mathrm{F}),(P^{\textrm{max}}_\mathrm{T}, P_\mathrm{M}|^{\{f_\mathrm{M}\}}, P^{\textrm{max}}_\mathrm{F}),$ \\ \\ $(P^{\textrm{max}}_\mathrm{T}, P^{\textrm{max}}_\mathrm{M}, P_\mathrm{F}|^{\{f_\mathrm{F}\}}), (P^{\textrm{max}}_T, P^{\textrm{max}}_\mathrm{M}, P^{\textrm{max}}_\mathrm{F} )$} \\ \hline
\multirow{10}{*}{\makecell{Edge point -\\ Two thresholds \\ satisfied}} &$ \gamma'_{\mathrm{E}} \leq \gamma^{\textrm{min}}_{E}$ & 4 & \makecell{\\$(P_\mathrm{T}|^{\{f_\mathrm{T}\}}, P^{\textrm{max}}_\mathrm{M}, P^{\textrm{max}}_\mathrm{F})$, $(P_\mathrm{T}|^{\{f_\mathrm{F}\}}, P^{\textrm{max}}_\mathrm{M}, P^{\textrm{max}}_\mathrm{F})$ \\ \\ $(P^{\textrm{max}}_\mathrm{T}, P_\mathrm{M}|^{\{f_\mathrm{M}\}}, P^{\textrm{max}}_\mathrm{F}), (P^{\textrm{max}}_\mathrm{T}, P_\mathrm{M}|^{\{f_\mathrm{F}\}}, P^{\textrm{max}}_\mathrm{F})$} \\ \cline{2-4}
&$ \gamma'_\mathrm{R} \leq \gamma^{\textrm{min}}_\mathrm{R}$ & 4 & \makecell{\\$(P^{\textrm{max}}_\mathrm{T}, P_\mathrm{M}|^{\{f_\mathrm{T}\}}, P^{\textrm{max}}_\mathrm{F})$, $(P^{\textrm{max}}_\mathrm{T}, P_\mathrm{M}|^{\{f_\mathrm{M}\}}, P^{\textrm{max}}_\mathrm{F})$ \\ \\ $(P^{\textrm{max}}_\mathrm{T}, P^{\textrm{max}}_\mathrm{M}, P_\mathrm{F}|^{\{f_\mathrm{T}\}})$, $(P^{\textrm{max}}_\mathrm{T}, P^{\textrm{max}}_\mathrm{M}, P_\mathrm{F}|^{\{f_\mathrm{F}\}})$}\\ \cline{2-4}
&$ \gamma'_\mathrm{C} \leq \gamma^{\textrm{min}}_\mathrm{C}$ & 4 & \makecell{\\$(P_\mathrm{T}|^{\{f_\mathrm{T}\}},P^{\textrm{max}}_\mathrm{M} , P^{\textrm{max}}_\mathrm{F}), (P_\mathrm{T}|^{\{f_\mathrm{M}\}},P^{\textrm{max}}_\mathrm{M} , P^{\textrm{max}}_\mathrm{F})$ \\ \\ $(P^{\textrm{max}}_\mathrm{T}, P^{\textrm{max}}_\mathrm{M}, P_\mathrm{F}|^{\{f_\mathrm{M}\}}), (P^{\textrm{max}}_\mathrm{T}, P^{\textrm{max}}_\mathrm{M}, P_\mathrm{F}|^{\{f_\mathrm{F}\}})$}\\ \hline
\multirow{5}{*}{\makecell{Edge point -\\ One threshold \\ satisfied}} &$ \gamma'_{\mathrm{E}} \leq \gamma^{\textrm{min}}_{\mathrm{E}} \text{ and } \gamma'_\mathrm{C} \leq \gamma^{\textrm{min}}_\mathrm{C}$ & 2 & \makecell{\\$(P_\mathrm{T}|^{\{f_\mathrm{M}\}}, P^{\textrm{max}}_\mathrm{M}, P^{\textrm{max}}_\mathrm{F}), (P_\mathrm{T}|^{\{f_\mathrm{F}\}}, P^{\textrm{max}}_\mathrm{M}, P^{\textrm{max}}_\mathrm{F})$} \\ \cline{2-4}
&$ \gamma'_\mathrm{R} \leq \gamma^{\textrm{min}}_\mathrm{R}\text{ and } \gamma'_\mathrm{E} \leq \gamma^{\textrm{min}}_{E}$ & 2 & \makecell{\\$(P^{\textrm{max}}_\mathrm{T}, P_\mathrm{M}|^{\{f_\mathrm{T}\}}, P^{\textrm{max}}_\mathrm{F}), (P^{\textrm{max}}_\mathrm{T}, P_\mathrm{M}|^{\{f_\mathrm{F}\}}, P^{\textrm{max}}_\mathrm{F})$}\\ \cline{2-4}
&$ \gamma'_\mathrm{R} \leq \gamma^{\textrm{min}}_\mathrm{R}\text{ and } \gamma'_\mathrm{C} \leq \gamma^{\textrm{min}}_\mathrm{C}$ & 2 & \makecell{\\$(P^{\textrm{max}}_\mathrm{T}, P^{\textrm{max}}_\mathrm{M}, P_\mathrm{F}|^{\{f_\mathrm{T}\}}), (P^{\textrm{max}}_\mathrm{T}, P^{\textrm{max}}_\mathrm{M}, P_\mathrm{F}|^{\{f_\mathrm{M}\}})$}\\ \hline
\end{tabular}
\end{center} \label{table:points}
\end{table*}

\subsection{Vertices of the Power Region}

We identify all the vertices of the power region for the different interference scenarios, from which one of the points will give us the suboptimal powers that maximize the sum rate. All the vertex points are summarized in Table \ref{table:points}. The notation $\{P_a,P_b\}|^{\{f_a,f_b\}}$ means solve for powers $P_a$ and $P_b$ using simultaneous equations $f_a$ and $f_b$ with the other power maximized, where $a,b \in\{\mathrm{T},\mathrm{M},\mathrm{F}\}$. How these are obtained is explained in detail below:

\textbf{\textit{Face points with one power maximized:}} There exists vertices that lie on a face of the cube and are formed from the intersection of two planes, e.g., point $F$ in Fig. \ref{fig:all_thresholds}. There are nine such vertices (three faces with three ways of choosing two intersecting planes for each face). These vertices can be found by solving two plane equations simultaneously with the power corresponding to the third face maximized. In general, it is difficult to identify exactly which of these nine points may be optimal for a given interference scenario. Thus, we need to test all nine vertices.

\textbf{\textit{Edge points with maximum powers satisfying all thresholds:}} Consider the case where the three planes are orthogonal, as shown in Fig. \ref{fig:all_thresholds}. In this case, the power region includes the top corner of the cube, where all three powers are maximized, and three other corner points where the planes intersect the edges of the cube, which we shall label as {\it edge points}. Since the top corner lies in the power region, this indicates that when all powers are maximized, all three SINRs $\gamma_{r}$ are greater than their minimum thresholds $\gamma^{\textrm{min}}_{r}$. For the rest of this section, we denote $\gamma'_{r}$ as the SINR for each node when all powers are at their maximum. There are four such points, as summarized in Table I. Note that the same SINR scenario can occur even when the planes are not perpendicular.\footnote{In fact, perpendicular planes which each only intersect one axis corresponds to an interference-free scenario.} The distinctive feature of this scenario is that the top corner is within the region spanned by the planes, and that each plane only intersects one of the maximum power edges of the cube.

\textbf{\textit{Edge points with maximum powers satisfying two thresholds:}} To visualize this scenario, imagine tilting the planes pivoted at $Q$ to form new power regions. For instance, if we tilt only the FAP plane upwards, it will eventually pass through the top corner and intersect the other two top edges. These two additional edge points ($B$ and $D$ in Fig. \ref{fig:two_thresholds}) add to the existing two edge points ($A$ and $\mathrm{C}$) to give a total of four edge points on the cube's edges. Since the top corner point will now be below the FAP plane, this means that $\gamma'_{\mathrm{E}} \leq \gamma^{\textrm{min}}_{\mathrm{E}}$. Similar arguments can be made for the other two planes, giving us three cases where there are a total of four corner points in the power region, each case corresponding to one $\gamma'_{r}$ that is less than its respective threshold.

\textbf{\textit{Edge points with maximum powers satisfying one threshold:}} For scenarios where two $\gamma'_{r}$ fail to reach their thresholds and only one is met, the two planes will be tilted such that the corner point lies outside both their feasible regions, as shown in Fig. \ref{fig:one_threshold} where the FAP and MBS planes lie above and to the left of the top corner respectively. In these cases, the feasible region will intersect one of the three corner edges at two points. Fig. \ref{fig:one_threshold} illustrates the maximum MBS power edge being intersected at two points $A$ and $B$ by the D2D and FAP planes respectively. Note that the MBS plane also intersects the same edge, but that point of intersection is outside the power region. Thus, we get two points, each corresponding to a set of conditions.

\section{Resource allocation in dedicated and cellular modes}\label{sec:resource}

If mode selection decides that the D2D pair can transmit using either dedicated or cellular mode, time and/or frequency resources must be allocated. We make the following assumptions for resource sharing in both dedicated D2D and cellular mode: (i) since cellular frequencies are used, there is a minimum rate guarantee for each user, including the DRx, (ii) there are enough resources to meet all users' minimum rate requirements, and (iii) at any one time, one transmitter can only operate in either uplink or downlink, i.e., half duplex. Consideration of full duplex transmitters~\cite{Wang2015_mag} is outside the scope of this work.

\subsection{Problem Formulation}
Since there is no interference in both dedicated and cellular modes and all powers can be maximized, the SINR at each receiver is the same as the signal-to-noise ratio (SNR) at that receiver, given as\footnote{With slight abuse of notation but for the sake of simplicity, we use the symbol $\gamma_r$ for SNR, where as in Section II we denoted minimum SINR at a receiver as $\gamma_r^{\mathrm{min}}$, where $r \in\{\mathrm{\mathrm{R}},\mathrm{\mathrm{C}},\mathrm{E}\}$ is the index for the receivers.}
\ifCLASSOPTIONonecolumn
\begin{eqnarray}
\gamma_\mathrm{C} &=& \frac{g_{\mathrm{M,C}}P^{\textrm{max}}_\mathrm{M}}{\sigma^2},\qquad \gamma_\mathrm{R} = \frac{g_{\mathrm{T},\mathrm{R}}P^{\textrm{max}}_\mathrm{T}}{\sigma^2},\qquad
\gamma_{\mathrm{E}} = \frac{g_{\mathrm{F,E}}P^{\textrm{max}}_\mathrm{F}}{\sigma^2}.
\end{eqnarray}
\else
\begin{subequations}
\begin{align}
\gamma_\mathrm{C} &= \frac{g_{\mathrm{M,C}}P^{\textrm{max}}_\mathrm{M}}{\sigma^2},\\
\gamma_\mathrm{R} &= \frac{g_{\mathrm{T},\mathrm{R}}P^{\textrm{max}}_\mathrm{T}}{\sigma^2},\\
\gamma_{\mathrm{E}} &= \frac{g_{\mathrm{F,E}}P^{\textrm{max}}_\mathrm{F}}{\sigma^2}.
\end{align}
\end{subequations}
\fi

We formulate a general optimization for a Long Term Evolution (LTE)-like resource grid with distinct resource blocks as follows
\begin{align}
\underset{B^i_r}{\text{maximize}} &\hspace{3mm}& \sum_r\sum_i^{N^t} B^i_r\delta^f\delta^t\log_2\left(1+\frac{\gamma_{r}}{B^i_r\delta^f}\right) \\
\text{subject to} &\hspace{3mm}& \sum_i^{N^t} B^i_r\delta^f\delta^t\log_2\left(1+\frac{\gamma_{r}}{B^i_r\delta^f}\right) \geq \mathcal{R}^{min}_r\label{eq:freqcond}
\end{align}where $B^i_r$ is the number of resource blocks for user $r$ at the $i$th time interval, $\delta^f_{r}$ and $\delta^t_{r}$ are the (constant) fractions representing the portion of each frequency and time block compared to the total grid respectively, and $N^t$ is the total number of time intervals for the resource grid. The total number of blocks allocated to user $r$ is therefore $\sum_i^{N^t} B^i_r$. Note that we divide the SNR by the frequency portion as we define $\sigma^2$ with respect to the entire bandwidth, resulting in equal noise power density [13].

The general formulation is difficult to solve, and in practice requires numerical methods. In order to gain insight into generalizations for arbitrary number of users and to obtain closed form solutions, we can show that for a given $\sum_i^{N^t} B^i_r = \mathcal{N}$ number of resource blocks and assuming that each block is no more preferable to any other, allocating resources across frequency will produce higher rates than allocating across time or in a random manner. For ease of proof and without loss of generality, we assume high SNR such that $\log_2 (1+SNR) \approx \log_2(SNR)$. Using $\log_2(A) + \log_2(B) = \log_2(AB)$ and the arithmetic-geometric mean inequality, which states that the maximum of a product of terms with a sum constraint occurs when all terms are equal, we find that the maximum of
\begin{align}
& \sum_i^{N^t} B^i_r \log_2\left(\frac{\gamma_{r}}{B^i_r}\right) =\log_2 \prod_i^{N^t} \left(\frac{\gamma_{r}}{B^i_r}\right)^{B^i_r} \\
& \text{subject to} \hspace{3mm}\sum_i^{N^t} B^i_r = \mathcal{N}
\end{align} occurs when all $B^i_r$ are equal. In other words, using equal bandwidth allocation across all time intervals for each user will provide the largest rates. Thus, although we can generalize resource allocation to be compatible with arbitrary resource block allocations, frequency allocation will give higher rates compared to other approaches for a given number of resources blocks.

Since frequency allocation can be solved in closed form, we analyze frequency allocation formulated as follows:
\begin{subequations}
\begin{eqnarray}
\underset{x_r}{\text{maximize}} &\hspace{3mm}& \sum x_{r}\log_2\left(1+\frac{\gamma_{r}}{x_{r}}\right)\\
\text{subject to} &\hspace{3mm}& x_{r}\log_2\left(1+\frac{\gamma_{r}}{x_{r}}\right) \geq \mathcal{R}^{min}_r\label{eq:freqcond}
\end{eqnarray}
\end{subequations}
where the factor $x_r$ is a function of the resource portions $0 \leq \alpha, \alpha', \beta, \beta' \leq 1$. For ease of analysis, we study only the case where an exact bandwidth is allocated across all time intervals. 

\begin{figure*}[t!]

\begin{subfigure}[t]{0.45\textwidth}
\includegraphics[width=1\textwidth]{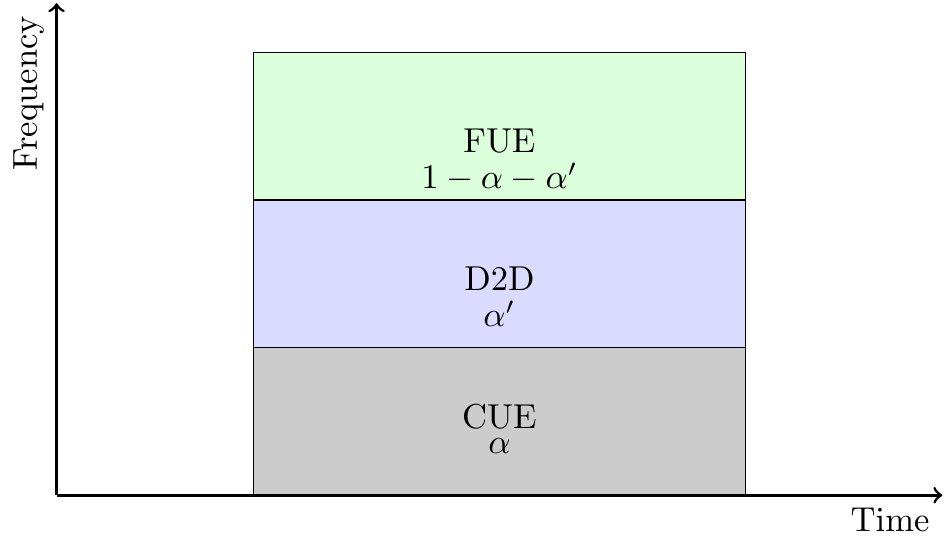}
\caption{Frequency sharing in dedicated mode.}
\label{fig:dedicated_freq}
\end{subfigure}
~~~~~~~~~~~~~~~~~~~~~
\begin{subfigure}[t]{0.45\textwidth}
\includegraphics[width=1\textwidth]{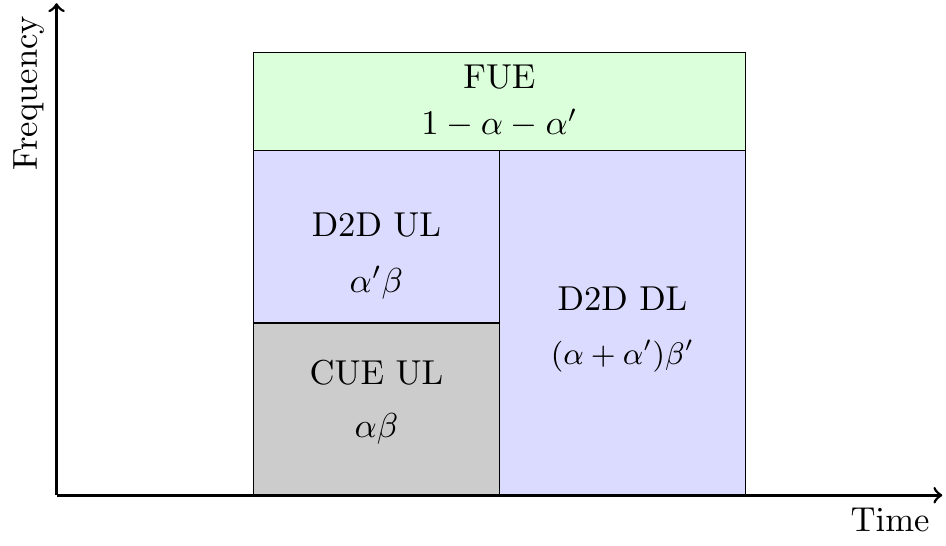}
\caption{Frequency sharing in cellular mode.}
\label{fig:cellular_freq}
\end{subfigure}

\caption{Illustration of resource allocation in dedicated and cellular modes.}
\end{figure*}

\subsection{Frequency sharing in dedicated D2D mode}
For the allocation structure illustrated in Fig. \ref{fig:dedicated_freq}, we want to maximize
\ifCLASSOPTIONonecolumn
\begin{align} \label{eq:dedicated_freq}
\mathcal{R}_{2} = \alpha \log_2\left(1+\frac{\gamma_\mathrm{C}}{\alpha}\right) + \alpha'\log_2\left(1+\frac{\gamma_\mathrm{R}}{\alpha'}\right) + (1-\alpha-\alpha')\log_2\left(1+\frac{\gamma_{\mathrm{E}}}{1-\alpha-\alpha'}\right).
\end{align}
\else
\begin{align} \label{eq:dedicated_freq} \nonumber
\mathcal{R}_{2} & = \alpha \log_2\left(1+\frac{\gamma_\mathrm{C}}{\alpha}\right) + \alpha'\log_2\left(1+\frac{\gamma_\mathrm{R}}{\alpha'}\right) \\
& + (1-\alpha-\alpha')\log_2\left(1+\frac{\gamma_{\mathrm{E}}}{1-\alpha-\alpha'}\right).
\end{align}
\fi

\subsubsection{Unconstrained}
With no minimum rate constraints, we differentiate (\ref{eq:dedicated_freq}) with respect to $\alpha$ and $\alpha'$, and simultaneously solve for $\frac{\partial \mathcal{R}_{2}}{\partial \alpha} =0$ and $\frac{\partial \mathcal{R}_{2}}{\partial \alpha'} =0$, which gives us the solutions
\begin{subequations}
\begin{align}
\alpha = \frac{\gamma_\mathrm{C}}{\gamma_\mathrm{C} + \gamma_\mathrm{R}+ \gamma_{\mathrm{E}}},\\ \alpha' = \frac{\gamma_\mathrm{R}}{\gamma_\mathrm{C} + \gamma_\mathrm{R} + \gamma_{\mathrm{E}}}.
\end{align}
\end{subequations}

Substituting the above into (\ref{eq:dedicated_freq}) and simplifying, the optimal sum rate is
\begin{equation}
\mathcal{R}_{2}^{\textrm{opt}}=\log_2(1+\gamma_\mathrm{C} + \gamma_\mathrm{R} + \gamma_{\mathrm{E}}).
\end{equation}

\subsubsection{Constrained}
To meet each user's minimum rate requirement, we require the solution to
\begin{equation} \label{eq:Lambert_W}
\alpha \log_2\left(1+\frac{\gamma_r}{\alpha}\right) = \mathcal{R}^{\textrm{min}}_{r}
\end{equation}

\noindent for each user. The solution can be written in terms of the Lambert W-function (see Appendix \ref{appendix:Lambert_W}), but there is no analytical solution that can be expressed using elementary functions. A simple numerical line search along $0 \leq \alpha \leq 1$ can be used to find an optimal solution.

\subsection{Frequency sharing in cellular D2D mode}
In frequency sharing cellular mode, because there is only one MBS transmitter, the D2D UL and CUE UL must occur at the same time, with D2D DL occurring immediately afterwards. The FUE can be allocated subbands at any time as it is served by a separate transmitter. Therefore, the allocation scheme follows the partitions as shown in Fig. \ref{fig:cellular_freq}.

In frequency sharing in cellular mode, the sum rate to be optimized is
\ifCLASSOPTIONonecolumn
\begin{align} \label{eq:cellular_freq}
\mathcal{R}_{4} &= \alpha \beta \log_2\left(1+\frac{\gamma_{\mathrm{M}, \mathrm{UL}}}{\alpha}\right) + \min\left(\alpha' \beta \log_2\left(1+\frac{\gamma_{\mathrm{R}, \mathrm{UL}}}{\alpha'}\right), (\alpha+\alpha') \beta' \log_2\left(1+\frac{\gamma_{\mathrm{R}, \mathrm{DL}}}{\alpha+\alpha'}\right)\right) \nonumber \\ & + (1-\alpha-\alpha')\log_2\left(1+\frac{\gamma_{\mathrm{E}}}{(1-\alpha-\alpha')}\right),
\end{align}
\else
\small
\begin{align} \label{eq:cellular_freq}
& \mathcal{R}_{4} = \alpha \beta \log_2\left(1+\frac{\gamma_{\mathrm{M}, \mathrm{UL}}}{\alpha}\right) \\ \nonumber & + \min\left(\alpha' \beta \log_2\left(1+\frac{\gamma_{\mathrm{R}, \mathrm{UL}}}{\alpha'}\right), (\alpha+\alpha') \beta' \log_2\left(1+\frac{\gamma_{\mathrm{R}, \mathrm{DL}}}{\alpha+\alpha'}\right)\right) \\ \nonumber & + (1-\alpha-\alpha')\log_2\left(1+\frac{\gamma_{\mathrm{E}}}{(1-\alpha-\alpha')}\right),
\end{align} \normalsize
\fi

\noindent where $\beta + \beta' = 1$\footnote{Setting $\beta' = 0$ would be equivalent to having two cellular users, and the solution would be the same as that in Section V-B.} and $\gamma_{r, \mathrm{UL}} = \frac{g_{\mathrm{C,M}}P^\mathrm{max}_\mathrm{C}}{\sigma^2}$ is the SNR at the MBS during CUE UL with the CUE transmitting at its maximum power $P^\mathrm{max}_\mathrm{C}$.

\subsubsection{Unconstrained}

We define uplink and downlink rates as
\begin{subequations}
\begin{align}
\mathcal{R}_{\mathrm{UL}} & = \log_2\left(1+\frac{\gamma_{\mathrm{R}, \mathrm{UL}}}{\alpha'}\right), \\
\mathcal{R}_{\mathrm{DL}} & = \log_2\left(1+\frac{\gamma_{\mathrm{R}, \mathrm{DL}}}{\alpha+\alpha'}\right).
\end{align}
\end{subequations}

To simplify (\ref{eq:cellular_freq}) into an expression involving only $\alpha$ and $\alpha'$, we note that the maximum sum rate occurs when $\alpha' \beta \mathcal{R}_\mathrm{UL} = (\alpha+\alpha') \beta' \mathcal{R}_{\mathrm{DL}}$, with the solution given by
\begin{equation}
\beta = \frac{\mathcal{R}_{\mathrm{DL}}}{\frac{\alpha'}{\alpha + \alpha'}\mathcal{R}_{\mathrm{UL}} + \mathcal{R}_{\mathrm{DL}}}.
\end{equation}

Substituting the above, the rate expression for D2D is
\begin{equation}
\mathcal{R}_d(\alpha) = \frac{\alpha' \mathcal{R}_{\mathrm{UL}}\mathcal{R}_{\mathrm{DL}}}{\frac{\alpha'}{\alpha + \alpha'}\mathcal{R}_{\mathrm{UL}} + \mathcal{R}_{\mathrm{DL}}}.
\end{equation}

Therefore, we can simplify (\ref{eq:cellular_freq}) to
\ifCLASSOPTIONonecolumn
\begin{align} \label{eq:cellular_freq_2}
\mathcal{R}_{4} = \frac{\alpha \mathcal{R}_{\mathrm{DL}} \log_2(1+\frac{\gamma_{\mathrm{R}, \mathrm{UL}}}{\alpha})}{\frac{\alpha'}{\alpha + \alpha'}\mathcal{R}_{\mathrm{UL}} + \mathcal{R}_{\mathrm{DL}}} + \mathcal{R}_d(\alpha) + (1-\alpha-\alpha')\log_2\left(1+\frac{\gamma_{\mathrm{E}}}{(1-\alpha-\alpha')}\right).
\end{align}
\else
\begin{align} \label{eq:cellular_freq_2}
\mathcal{R}_{4} & = \frac{\alpha \mathcal{R}_{\mathrm{DL}} \log_2(1+\frac{\gamma_{\mathrm{R}, \mathrm{UL}}}{\alpha})}{\frac{\alpha'}{\alpha + \alpha'}\mathcal{R}_{\mathrm{UL}} + \mathcal{R}_{\mathrm{DL}}} + \mathcal{R}_d(\alpha) \\ \nonumber &+ (1-\alpha-\alpha')\log_2\left(1+\frac{\gamma_{\mathrm{E}}}{(1-\alpha-\alpha')}\right).
\end{align}
\fi

\noindent A numerical search for $0 \leq\alpha + \alpha' \leq 1$ can be performed to maximize (\ref{eq:cellular_freq_2}).

\subsubsection{Constrained}
In this scenario, we desire to maximize (\ref{eq:cellular_freq_2}) under a minimum rate constraint for each user. Again, a numerical search can be performed to find the maximum.

\section{\Tr{Results and Discussion}}\label{simulation}

In this section, we present simulation results to illustrate the benefits of using our decision making framework over conventional cellular transmission for a potential D2D pair. Unless stated otherwise, simulation parameters presented in Table~\ref{tb:simpara} are used, which are similar to those adopted in \cite{Yu2014}. We use $(x,y)$ coordinates in meters to describe node locations.
\begin{table}[h!]
\normalsize
\caption{Values of Simulation Parameters}
\centering
\begin{tabular}{ | p{4cm} | p{4cm} |}
\hline
\bf{Parameter} & \bf{Value} \\ \hhline{|=|=|}
Bandwidth & $20$ MHz \\ \hline
Noise spectral density & $-174$ dBm/Hz \\ \hline
Max MBS transmit power & $P^{\textrm{max}}_\mathrm{M} = 43$ dBm \\ \hline
Max FAP transmit power & $P^{\textrm{max}}_\mathrm{F} = 21$ dBm \\ \hline
Max DTx transmit power & $P^{\textrm{max}}_\mathrm{T} = 23$ dBm \\ \hline
DTx coordinates & Varying along $x=y$ \\ \hline
DRx coordinates & Varying along $x=y$\\ \hline
MBS coordinates & (0, 0) \\ \hline
CUE coordinates & (500, 0) \\ \hline
FAP coordinates & (100, 200) \\ \hline
FUE coordinates & (110, 200) \\ \hline
DTx to DRx pathloss & $28$ + $40$log$_{10}(d)$ (dB) \\ \hline
MBS to CUE pathloss & $15.3$ + $37.6$log$_{10}(d_{\mathrm{M,C}})$ (dB)\\ \hline
FAP to FUE pathloss & $38.5$ + $20$log$_{10}(d_{F,E})$ (dB)\\ \hline
CUE minimum SINR & $\gamma^{\textrm{min}}_\mathrm{C} = 0$ dB \\ \hline
FUE minimum SINR & $\gamma^{\textrm{min}}_{\mathrm{E}} = 7$ dB \\ \hline
DRx minimum SINR & $\gamma^{\textrm{min}}_\mathrm{R} = 3$ dB \\ \hline
\end{tabular}\label{tb:simpara}
\end{table}
\begin{figure}
         \centering
    \includegraphics[width=0.48\textwidth,viewport=98 265 485 565,clip]{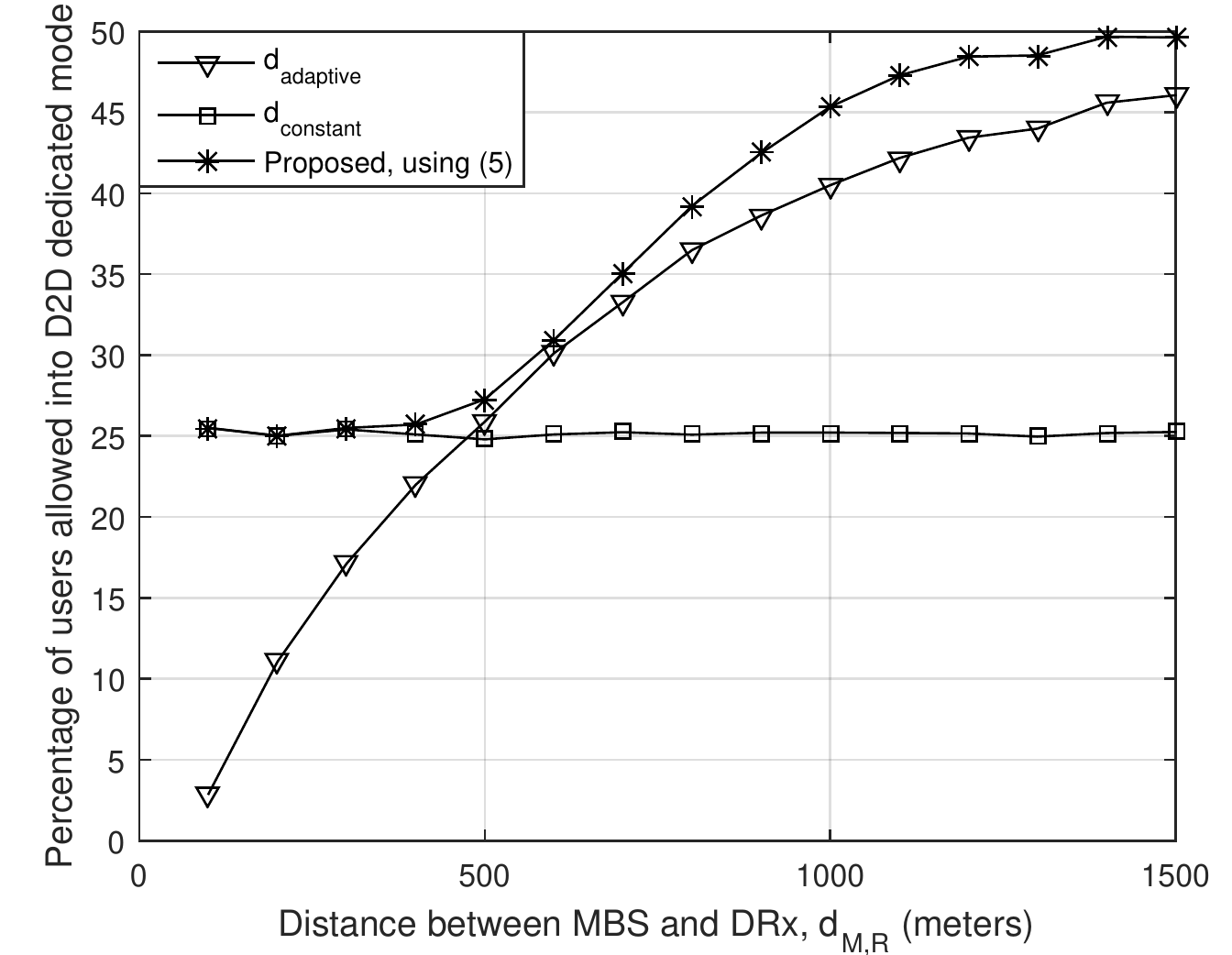}
    \caption{Percentage of potential D2D pairs entering dedicated mode. Predetermined threshold is better when interference is large, while adaptive threshold is better when interference is small.}
    \label{fig:new_framework}
\end{figure}
\begin{figure*}[t!]
\centering
\begin{subfigure}[t]{0.48\textwidth}
\centering
\includegraphics[width=\textwidth,viewport=98 270 485 565,clip]{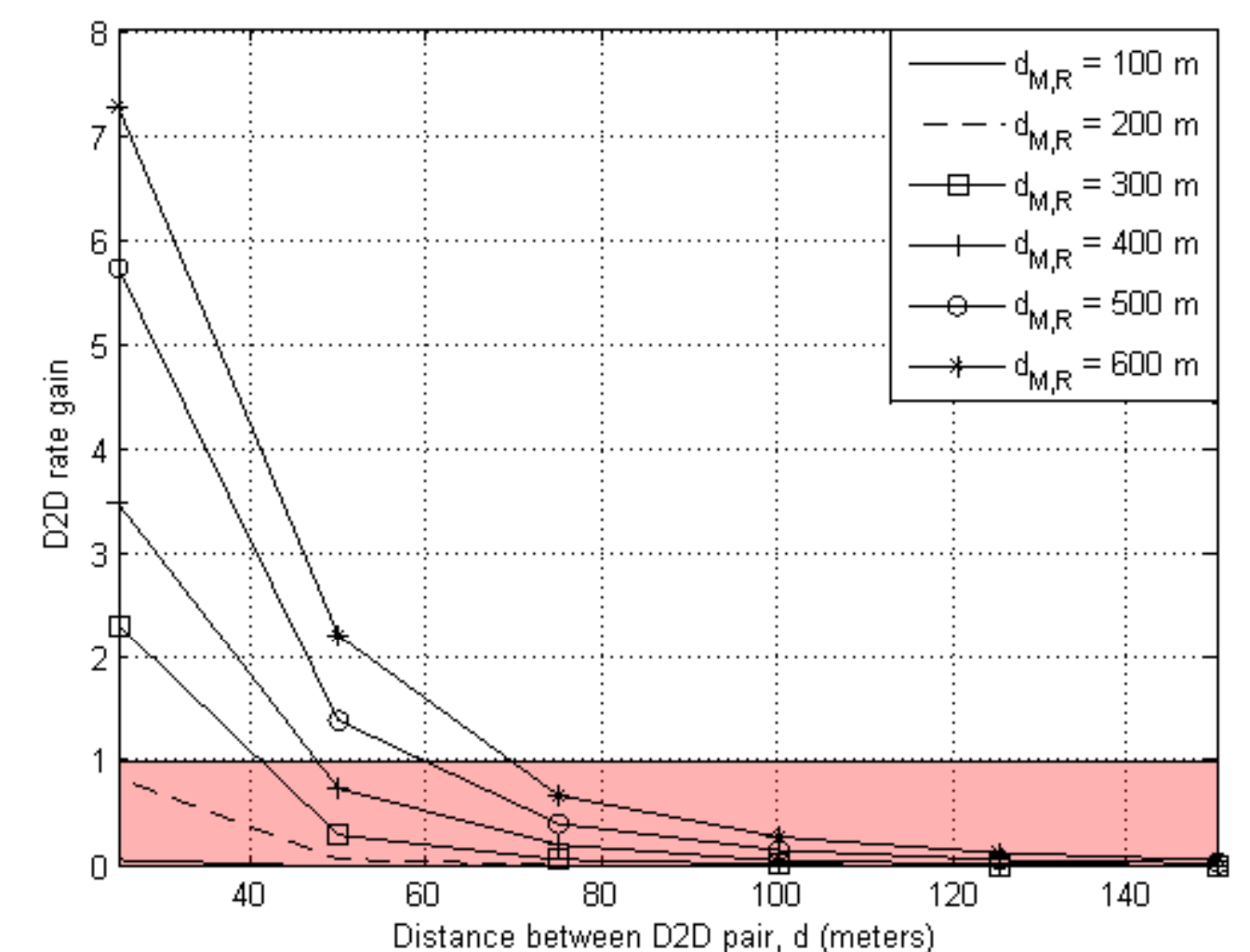}
\caption{D2D rate gain versus the distance between the DTx and DRx, $d$ for different MBS-DRx distance, $d_{\mathrm{M,R}}$. The shaded area below D2D rate gain of 1 represents the region where selecting D2D mode would be an incorrect decision.}
\label{fig:mode_selection_D2D}
\end{subfigure}
~
\begin{subfigure}[t]{0.48\textwidth}
\includegraphics[width=\textwidth,viewport=98 265 485 565,clip]{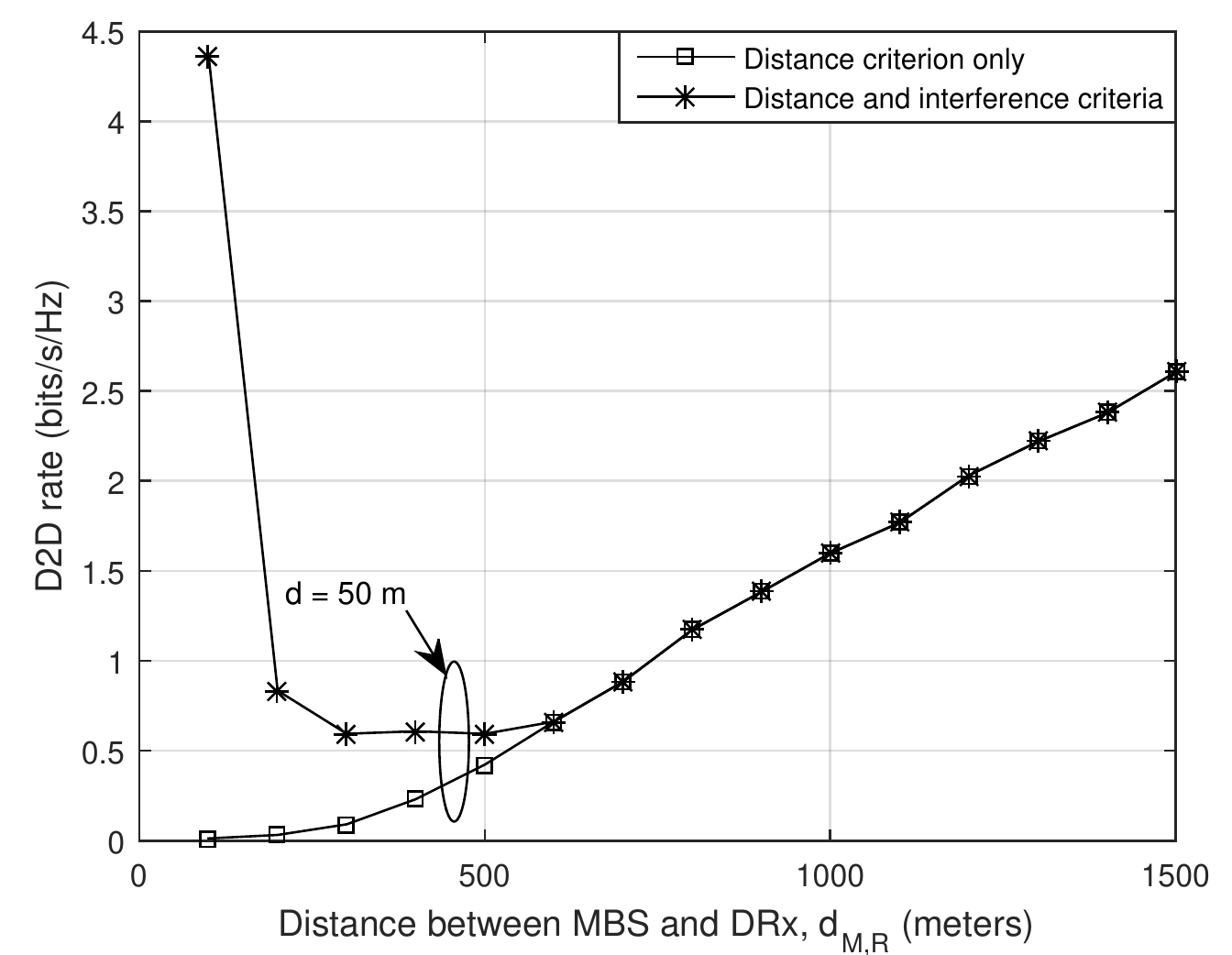}
\caption{D2D rate versus the distance between the MBS and DRx, $d_{\mathrm{M,R}}$, for mode selection using distance only criterion and two stage criteria.}
\label{fig:mode_selection_2_criteria}
\end{subfigure}
\caption{Performance of proposed D2D mode selection criteria.}
\end{figure*}
\subsection{Mode Selection}\label{results:modeselec}

We first show the advantages of using \eqref{eq:dist_new} compared to using a constant and adaptive distance threshold only. Setting $d_\textrm{constant} = 50$m and assuming orthogonal resources are available 50\% of the time, Fig. \ref{fig:new_framework} shows that picking the largest threshold between the predetermined and calculated gives the highest percentages of users entering dedicated mode. \Tr{When interference to the DRx is large, choosing a predetermined distance threshold is more beneficial. When the DRx is farther from an interference source, an adaptive threshold is the better choice as larger D2D separation distances can be tolerated.} It is evident that the proposed method captures the best features of the other two, and in fact slightly outperforms the best of both at every location tested.

When orthogonal resources are not available, Fig.~\ref{fig:mode_selection_D2D} plots the D2D rate gain versus the distance between the DTx and DRx, $d$. The D2D rate gain refers to the ratio between the D2D rate and the cellular rate, both under the same interference conditions. We can see that the D2D gain decreases when the D2D pair become farther apart and also when the DRx is closer to the MBS. This is in line with the discussion in Section \ref{framework_mode} since: (i) when the DRx is closer to the MBS (the largest interference source), using cellular mode should provide higher rates than an incorrect D2D mode decision since there would have been more interference, and (ii) when the D2D pair separation distance increases, cellular mode should provide higher rates since D2D mode would be weaker with increasing separation distance under constant transmit power. In Fig.~\ref{fig:mode_selection_D2D}, the D2D separation distance at which each curve intersects the boundary of this region can be calculated using (\ref{eq:distance_threshold}). Our calculated and simulated values were found to be in close agreement. For example, the calculated separation distance for $d_\mathrm{M,R} = 600$ m is $75.9$ m, while the simulations give a value of $71$ m.

Fig. \ref{fig:mode_selection_2_criteria} shows the actual rates experienced by a DRx when using the proposed mode selection method satisfying \eqref{eq:MS} and when using just the D2D minimum distance criterion with $d=50$ m. If D2D mode is always allowed for $d = 50$ m, the DRx can experience a smaller rate due to its close proximity to an MBS (or other large interference source), while using our proposed method will avoid such instances.

It is important to note that our results in this subsection do not suggest that cellular mode is superior to D2D mode. Rather, our results highlight that under some conditions, using a single criterion to determine mode selection can lead to an incorrect decision. We will show in Section \ref{orthogonal_simulation} that if D2D is operating in dedicated mode, it does in fact outperform regular cellular mode.

\subsection{Reuse Mode}

\ifCLASSOPTIONtwocolumn
\begin{figure}
    \includegraphics[width=0.48\textwidth,viewport=98 265 485 565,clip]{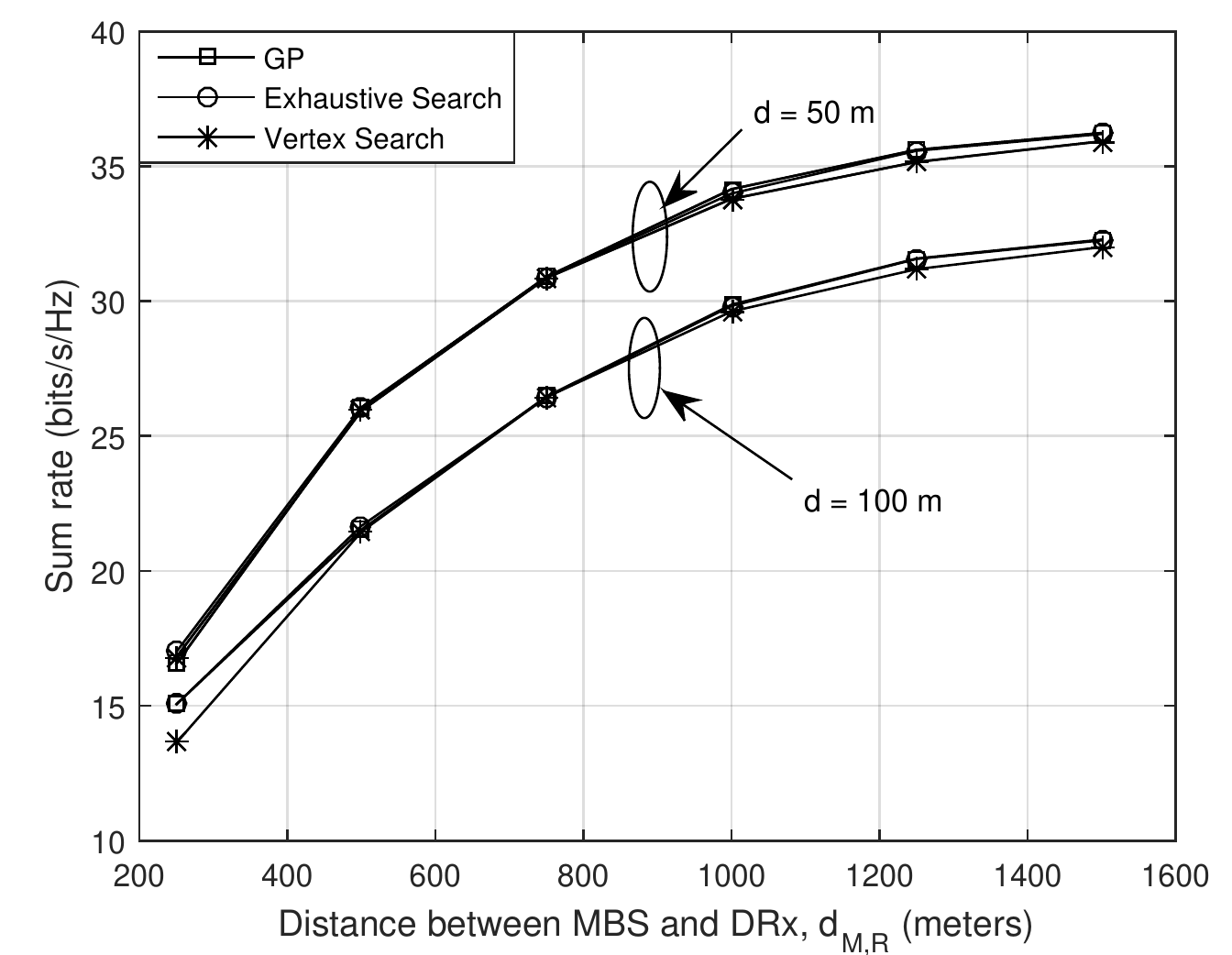}
    \caption{Sum rate in reuse mode with transmit powers determined using proposed near-optimal vertex search approach, geometric programming and exhaustive search.}
    \label{fig:reuse_mode}
     \end{figure}

     \begin{figure}
    \includegraphics[width=0.48\textwidth,viewport=98 265 485 565,clip]{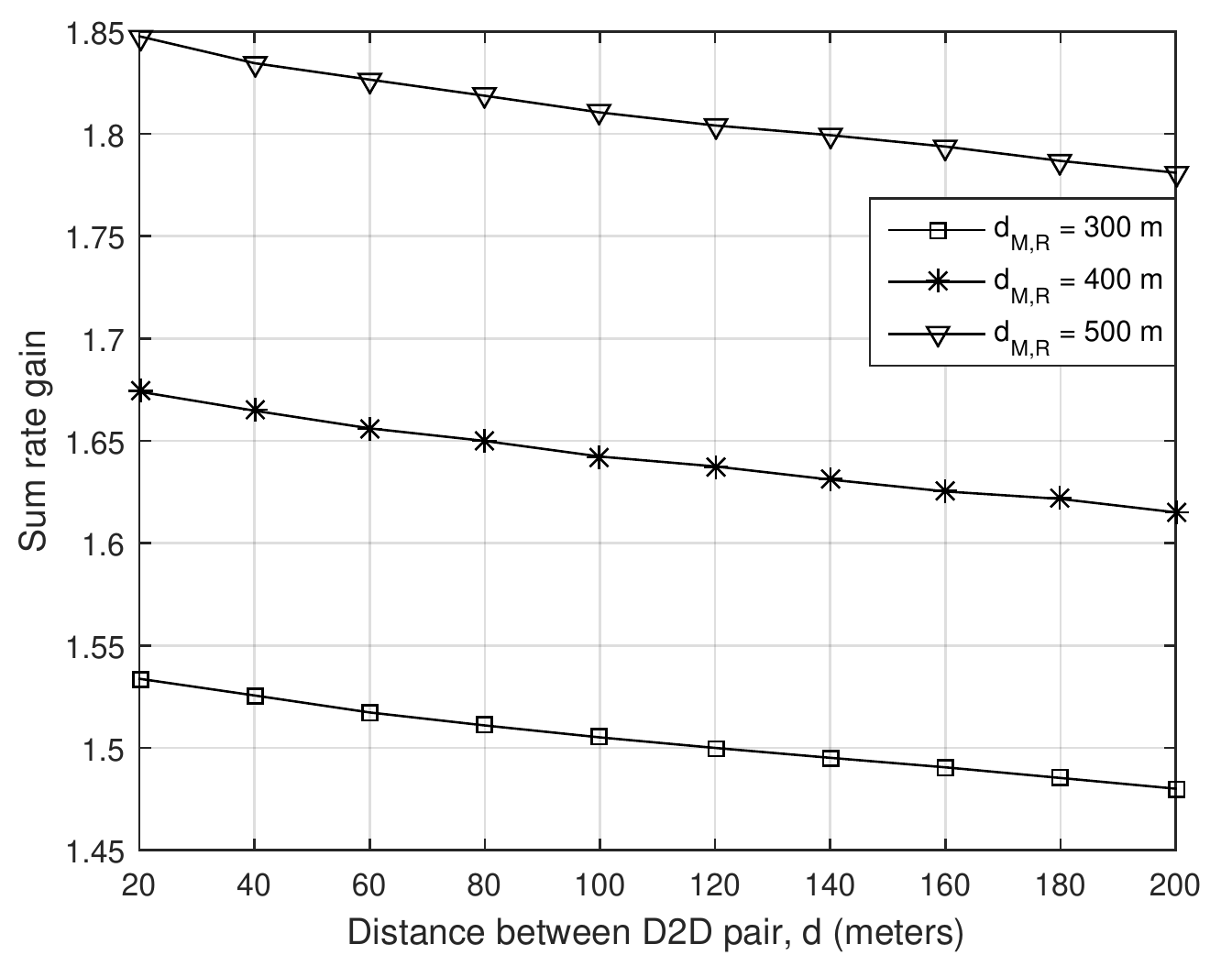}
    \caption{Sum rate gain versus the distance between the DTx and DRx, $d$ for constrained frequency resource sharing.}
    \label{fig:orthogonal_constrained}
\end{figure}
\fi
Fig.~\ref{fig:reuse_mode} plots the sum rate in reuse mode versus varying MBS-DRx distance, $d_{\mathrm{M,R}}$, comparing the near-optimal powers found using the proposed approach with those obtained from geometric programming (GP)~\cite{Chiang2007} and exhaustive search. We believe that using GP and exhaustive search serve as sufficient benchmarks - GP is one of the most common numerical approaches to finding near optimal solutions for the power control problem, while exhaustive search with a sufficiently fine step size confirms the optimality.

Our results show that for the considered parameters, our proposed method of searching the vertices of the power region and using the one that gives the maximum sum rate is comparable to optimal solutions. However, our method requires far fewer calculations than exhaustive search and GP, the latter of which relies on successive approximations with no prior indication on how many iterations are required. For example, using an Intel i7 3.2 GHz CPU with 16 GB RAM, for $d_{\mathrm{M,R}}=1000$ m in Fig. 7 GP took up to $44.5$ seconds to calculate a solution, exhaustive search took $49.8$ seconds, while our vertex search took only $1.2$ seconds, i.e., an improvement of around $40$ times over both benchmarks. Thus, GP can be unreliable in determining a suboptimal solution in sufficient time, while our vertex search approach will always return a suboptimal solution if the problem is feasible for small numbers of reuse powers. A further advantage of vertex search is that it always takes approximately the same time to calculate a solution for each realization, while the run time and accuracy of GP heavily depends on stoppage parameters.
\ifCLASSOPTIONonecolumn
\begin{figure*}
     \centering
     \begin{minipage}[t]{0.48\textwidth}
    \centering
    \includegraphics[width=1\textwidth,viewport=98 265 485 565,clip]{reuse_mode}
    \caption{Sum rate in reuse mode with transmit powers determined using proposed near-optimal vertex search approach, geometric programming and exhaustive search.}
    \label{fig:reuse_mode}
     \end{minipage}
     \centering
     \begin{minipage}[t]{0.48\textwidth}
         \centering
    \includegraphics[width=1\textwidth,viewport=98 265 485 565,clip]{orthogonal_constrained}
    \caption{Sum rate in reuse mode with transmit powers determined using proposed near-optimal vertex search approach, geometric programming and exhaustive search.}
    \label{fig:orthogonal_constrained}
     \end{minipage}
\end{figure*}
\fi
\subsection{Dedicated and Cellular Modes} \label{orthogonal_simulation}
Fig. \ref{fig:orthogonal_constrained} shows the sum rate gain, i.e., sum rate in dedicated mode divided by sum rate in cellular mode, under minimum rate requirements for each user. It is clear that dedicated D2D mode provides a greater sum rate when the D2D separation distance is small, and/or when the distance between the MBS and DRx is large.

It must be noted that the unconstrained dedicated and cellular modes offered similar sum rates under the simulation parameters, and thus their results are not shown. It is clear however that unconstrained dedicated sum rates will never be lower than their cellular counterparts since the D2D option is intended to improve overall system performance, and will not degrade the best performing user. Thus, we can conclude that D2D mode is more advantageous when users have individual rate constraints.

\subsection{\Tr{Scability Discussion}}
\Tr{Although we have presented our methodology using a simple system model, we can analyze the scalability with respect to increasing number of base stations and users. For our mode selection framework, the number of decisions scales linearly with the number of potential D2D pairs, and not the total number of users or base stations.
 
For reuse mode, we have presented three transmit powers to be optimized, leading to a 3-dimensional problem. Increasing the number of transmit powers increases the dimensionality of the problem, while increasing the number of users increases the number of planes, and further restricts the size of the power region. For $N$ powers, the power constraints form an $N$-dimensional hypercube, while the minimum SINR constraints further bound the region to form an $N$-dimensional polytope. Depending on which scenario the network is in (i.e., number of thresholds satisfied by max powers), the complexity of vertex search could increase exponentially at worst (e.g., in Fig. 3a), and linearly at best (Fig. 3c). However, if a small number of users share the same resource, since our vertex search avoids iterations, it can still be a more effective solution. Exact expressions for the vertices for $N$-dimensions is an interesting topic for future research.

For frequency sharing in dedicated mode (illustrated in Fig.~\ref{fig:dedicated_freq}), we prove in Appendix~\ref{appendix:freq_sharing_general} that the unconstrained case has a general solution for any number of transmitters and partitions. The general constrained case also has a solution given by solving \eqref{eq:Lambert_W} for each user. With increasing number of nodes (base station or user), the complexity would scale linearly as each additional node would require one additional equation to solve for (from differentiating \eqref{eq:dedicated_freq} or solving \eqref{eq:Lambert_W}). Frequency sharing in cellular mode (illustrated in Fig.~\ref{fig:cellular_freq}) can have various allocation structures due to the simultaneous uplink condition, and thus cannot be generalized in the present manner.
}

\section{Conclusion}\label{sec:conc}
We have presented a comprehensive mode selection, power control and resource allocation framework for D2D communication underlaying a two-tier cellular network.
Our proposed mode selection scheme allows D2D communications under stricter conditions, leading to more correct decision making and a higher rate of allowing dedicated mode. We have also proposed a geometric approach to determine near-optimal powers for power allocation in reuse mode with faster computational time than benchmark methods, and provided closed-form resource allocations for orthogonal D2D mode for any number of users.

There are numerous interesting additional features and directions for future research in our work. Energy efficiency could be used instead of sum rate as an objective, which would be particularly relevant for uplink scenarios, while imperfect CSI is a practical issue that can also be considered.
\appendices
\section{Proof of Proposition~\ref{prop:1}}\label{appendix:SINR_quasiconvex}

From \cite{Boyd2004}, a differentiable function is quasiconvex if and only if
\ifCLASSOPTIONonecolumn
\begin{align} \label{eq:quasi_def}
& f(y_1, \hdots, y_n) \leq f(x_1, \hdots, x_n) \Rightarrow \nabla f(x_1, \hdots, x_n)^T(y_1-x_1,\hdots,y_n-x_n)^T \leq 0
\end{align}
\else
\begin{align} \label{eq:quasi_def}
& f(y_1, \hdots, y_n) \leq f(x_1, \hdots, x_n) \\
& \Rightarrow \nabla f(x_1, \hdots, x_n)^T(y_1-x_1,\hdots,y_n-x_n)^T \leq 0
\end{align}
\fi
Although we can apply this to the sum SINR function directly, we note that since the addition of bounds and differentiation are preserved under addition, it is sufficient to show that each SINR is quasiconvex in order to show that sum SINR is quasiconvex\footnote{In general, the sum of quasi-convex functions may not be quasi-convex.}. Further, we can ignore the noise constant in the denominator as it does not change the convexity behaviour or shape of each fraction.

Consider the generic definition of sum SINR
\begin{equation}\label{eq:sum_SINR}
S = \sum_{i=1}^{N}\frac{P_i}{a_i}
\end{equation}

\noindent where $a_i = \sum_{j \neq i} P_j + \sigma^2$. For $N$ varying powers, if the numerator of an SINR fraction is constant, i.e., a power that is not varying, then
\begin{equation}
\frac{k}{P_1 + \hdots + P_N}
\end{equation} for varying powers $P_1, \hdots, P_N$ and constant $k$ is clearly quasiconvex as it follows a hyperbolic shape. If the numerator is a varying power, i.e.,
\begin{equation}
\frac{P_1}{P_2 + \hdots + P_{N-1}},
\end{equation} then using the second inequality in (\ref{eq:quasi_def}) we find that for two sets of powers $\{P_1, \hdots, P_N\}$ and $\{P'_1, \hdots, P'_N\}$,
\ifCLASSOPTIONonecolumn
\begin{align} \nonumber
\nabla S(P_1, \hdots, P_n)^T(P'_1-P_1,\hdots,P'_n-P_n)^T &= \begin{pmatrix}
\frac{1}{P_2+\hdots+ P_N} & \hdots & \frac{-P_1}{(P_2+\hdots+ P_N)^2}
\end{pmatrix}\begin{pmatrix}
P'_1-P_1\\
\vdots\\
P'_N-P_N
\end{pmatrix}=\\ \nonumber
\frac{P_1'-P_1}{P_2+\hdots+ P_N} - P_1\sum_{i=2}^{N}\frac{P_i'-P_i}{(P_2+\hdots+ P_N)^2} &\leq 0,\\ \nonumber
(P_1'-P_1)(P_2+\hdots+ P_N) - P_1\sum_{i=2}^{N}(P_i'-P_i) &\leq 0, \\ \nonumber
P_1' (P_2+\hdots+ P_N)&\leq P_1(P_2'+\hdots+ P_N'), \\
\frac{P_1'}{P_2'+\hdots+ P_N'} &\leq \frac{P_1}{P_2+\hdots+ P_N}.
\end{align}
\else
\begin{align} \nonumber
& \nabla S(P_1, \hdots, P_n)^T(P'_1-P_1,\hdots,P'_n-P_n)^T = \\ \nonumber
& \begin{pmatrix}
\frac{1}{P_2+\hdots+ P_N} & \hdots & \frac{-P_1}{(P_2+\hdots+ P_N)^2}
\end{pmatrix}\begin{pmatrix}
P'_1-P_1\\
\vdots\\
P'_N-P_N
\end{pmatrix}=\\ \nonumber
& \frac{P_1'-P_1}{P_2+\hdots+ P_N} - P_1\sum_{i=2}^{N}\frac{P_i'-P_i}{(P_2+\hdots+ P_N)^2} \leq 0,\\ \nonumber
& (P_1'-P_1)(P_2+\hdots+ P_N) - P_1\sum_{i=2}^{N}(P_i'-P_i) \leq 0, \\ \nonumber
& P_1' (P_2+\hdots+ P_N) \leq P_1(P_2'+\hdots+ P_N'), \\
& \frac{P_1'}{P_2'+\hdots+ P_N'} \leq \frac{P_1}{P_2+\hdots+ P_N}.
\end{align}
\fi

\noindent This is the first inequality in (\ref{eq:quasi_def}) when $P_i' = y_i$, $P_i = x_i$. Thus, for any combination of varying powers, we find that sum SINR is quasiconvex.

\section{Proof of Proposition~\ref{prop:2}}\label{appendix:asymptotic}

Suppose we differentiate $S$ in \eqref{eq:sum_SINR} with respect to the most dominant power $P_i$:
\begin{equation}
\frac{d S}{d P_i} = \frac{1}{a_i} - \sum_{j \neq i}^{N} \frac{P_j}{a_j^2}.
\end{equation} If $P_i$ was a dominant power, we observe that the derivative will approach $1/a_i$ since $a_j \rightarrow \infty$ as $P_i \rightarrow \infty$.

Similarly, for the generic definition of sum rate
\ifCLASSOPTIONonecolumn
\begin{align} \label{eq:sum_rate}
\mathcal{R} = \sum_{i=1}^{N}\log_2\left(1+\frac{P_i}{\sum_{j\neq i}P_j + \sigma^2}\right) = \log_2\prod_{i=1}^{N}\left(1+\frac{P_i}{\sum_{j\neq i}P_j + \sigma^2}\right),
\end{align}
\else
\begin{align} \nonumber
\mathcal{R} & = \sum_{i=1}^{N}\log_2\left(1+\frac{P_i}{\sum_{j\neq i}P_j + \sigma^2}\right) \\ \label{eq:sum_rate}
& = \log_2\prod_{i=1}^{N}\left(1+\frac{P_i}{\sum_{j\neq i}P_j + \sigma^2}\right),
\end{align}
\fi
if we expand out the brackets in (\ref{eq:sum_rate}) ignoring the logarithm to obtain
\ifCLASSOPTIONonecolumn
\begin{align} \nonumber \label{eq:sum_rate_expand}
& \prod_{i=1}^{N}\left(1+\frac{P_i}{a_i}\right) = 1 + \sum_{i=1}^{N} \frac{P_i}{a_i}
+ \sum (\text{Products of SINRs two at a time}) \\ + &\sum (\text{Products of SINRs three at a time}) + \hdots + \prod_{i=1}^N\frac{P_i}{a_i}
\end{align}
\else
\begin{align} \nonumber \label{eq:sum_rate_expand}
& \prod_{i=1}^{N}\left(1+\frac{P_i}{a_i}\right) = 1 + \sum_{i=1}^{N} \frac{P_i}{a_i} \\ \nonumber
& + \sum (\text{Products of SINRs two at a time}) \\ \nonumber
& + \sum (\text{Products of SINRs three at a time}) \\
& + \hdots + \prod_{i=1}^N\frac{P_i}{a_i}
\end{align}
\fi
and differentiate with respect to $P_i$, we find that all the derivatives of the products of SINRs will contain $a_j^2$ in the denominator, and will approach $0$ as $P_i \rightarrow \infty$. Thus, both sum SINR and (\ref{eq:sum_rate_expand}) have the same {\it asymptotic gradient} of $1/a_i$ when one power dominates. Note that if we differentiate with respect to $P_i$, but $P_i$ was not the dominate power, both expressions will instead approach $-P_j/a_j^2$ if $P_j$ was the dominant power.

\section{Closed form solution for constrained frequency sharing in dedicated D2D mode}\label{appendix:Lambert_W}
To solve (\ref{eq:Lambert_W}), we need to manipulate (\ref{eq:Lambert_W}) to a form where we can use the Lambert W function. Firstly, we can rearrange and then exponentiate (\ref{eq:Lambert_W}) to get
\begin{equation}
\ln\left(1+\frac {\gamma_r}{\alpha}\right)=e^{ \mathcal{R}^\mathrm{min}_r\ln 2/\alpha}.
\end{equation}
Next, we need to introduce additional terms such that the exponent contains the left hand side, i.e.,
\begin{equation}
\ln\left(1+\frac {\gamma_r}{\alpha}\right)=2^{- \mathcal{R}^\mathrm{min}_r/\gamma_r}e^{\large\frac{ \mathcal{R}^\mathrm{min}_r\ln2}{\gamma_r}\left(1+\frac{\gamma_r}{\alpha}\right)}.
\end{equation}
Moving the exponential to the left hand side gives
\ifCLASSOPTIONonecolumn
\begin{align}
-\frac{ \mathcal{R}^\mathrm{min}_r\ln2}{\gamma_r}\ln\left(1+\frac {\gamma_r}{\alpha}\right)e^{-\large\frac{ \mathcal{R}^\mathrm{min}_r\ln2}{\gamma_r}\left(1+\frac{\gamma_r}{\alpha}\right)}
= -\frac{ \mathcal{R}^\mathrm{min}_r\ln2}{\gamma_r}2^{- \mathcal{R}^\mathrm{min}_r/{\gamma_r}}.
\end{align}
\else
\begin{align}
& -\frac{ \mathcal{R}^\mathrm{min}_r\ln2}{\gamma_r}\ln\left(1+\frac {\gamma_r}{\alpha}\right)e^{-\large\frac{ \mathcal{R}^\mathrm{min}_r\ln2}{\gamma_r}\left(1+\frac{\gamma_r}{\alpha}\right)} \\ \nonumber
& = -\frac{ \mathcal{R}^\mathrm{min}_r\ln2}{\gamma_r}2^{- \mathcal{R}^\mathrm{min}_r/\gamma_r}.
\end{align}
\fi
We can now apply the Lambert W function since the exponential is in the form $Ae^A$:
\begin{equation}
-\frac{ \mathcal{R}^\mathrm{min}_r\ln2}{\gamma_r}\ln\left(1+\frac {\gamma_r}{\alpha}\right) = \mathrm{W}\left(-\frac{ \mathcal{R}^\mathrm{min}_r\ln2}{\gamma_r}2^{- \mathcal{R}^\mathrm{min}_r/\gamma_r} \right).
\end{equation}
Rearranging for $\alpha$ gives the solution
\begin{equation}
\alpha = \frac{-\gamma_r \mathcal{R}^\mathrm{min}_r\ln2}{ \mathcal{R}^\mathrm{min}_r \ln2 + \gamma_r\mathrm{W}\left(-\frac{ \mathcal{R}^\mathrm{min}_r\ln2}{\gamma_r}2^{- \mathcal{R}^\mathrm{min}_r/\gamma_r} \right)}.
\end{equation} To ensure a real solution, we use the $-1$ branch of the Lambert W function.

\section{General solution for unconstrained frequency sharing in dedicated mode}\label{appendix:freq_sharing_general}
For $N$ transmitters, and hence $N$ partitions, sum rate is
\begin{equation} \label{eq:sum_rate_freq_general}
\mathcal{R} = \sum_{i=1}^{N}\alpha_i\log_2\left(1+\frac{\gamma_i}{\alpha_i}\right),
\end{equation} where $\sum \alpha_i = 1$ is the partition fraction and $\gamma_i$ is the SNR of each receiver.

In order to greedily maximize $\mathrm{R}$, we need to simultaneously solve the partial derivatives with respect to each $\alpha_i$, i.e. $\frac{\partial \mathcal{R}}{\partial \alpha_i} = 0$. This will give the relations
\begin{equation}\label{eq:alpha_gamma}
\frac{\gamma_i}{\alpha_i} = \frac{\gamma_k}{\alpha_k}
\end{equation} for $i,k = 1,\hdots,N$.
Setting $k=m$, and noting that $\alpha_m = 1-\sum_{k=1}^{N-1}\alpha_k$, we can rearrange (\ref{eq:alpha_gamma}) to obtain
\begin{equation}
\alpha_i = \left( 1-\sum_{k=1}^{N-1}\frac{\alpha_i \gamma_k}{\gamma_i}\right)\frac{\gamma_i}{\gamma_n}
=\frac{\gamma_i}{\gamma_m} - \frac{\alpha_i}{\gamma_m}\sum_{k=1}^{N-1}\gamma_k,
\end{equation} which can be simplified to
\begin{equation}
\alpha_i = \frac{\gamma_i}{\gamma_m + \sum_{k=1}^{N-1}\gamma_k} = \frac{\gamma_i}{\sum_{k=1}^{N}\gamma_k}.
\end{equation}
Thus, each resource partition fraction is equal to the fraction of the particular SNR over the total SNR. Substituting the above into (\ref{eq:sum_rate_freq_general}) will always give the maximum sum rate
\begin{equation}
\mathcal{R} = \log_2\left( 1+ \sum_{i=1}^N \gamma_i\right).
\end{equation}


\end{document}